\documentclass[twocolumn, showkeys, jpb,superscriptaddress,floatfix,nofootinbib]{revtex4-2}
\usepackage{graphicx} 

\usepackage[a4paper, top=2cm, bottom=2cm, left=2cm, right=2cm]{geometry}

\usepackage{xcolor}
\usepackage{comment}
\usepackage[normalem]{ulem}
\usepackage{hyperref}
\usepackage{amsmath}

\newcommand{\Gr}{$T_\mathrm{osc}/T_\mathrm{grow}$}

\begin{document}
\title{Flow–wave coupling synchronizes oscillations in growing active matter}

\author{Lara Koehler}
\email{These authors contributed equally to this work}
\affiliation{Cluster of Excellence Physics of Life, TU Dresden, Dresden, 01307 Germany}
\affiliation{Max Planck Institute for the Physics of Complex Systems, Nöthnitzer Straße 38, 01187 Dresden, Germany}
\author{Elissavet Sandaltzopoulou}
\email{These authors contributed equally to this work}
\affiliation{Cluster of Excellence Physics of Life, TU Dresden, Dresden, 01307 Germany}
\affiliation{Max Planck Institute of Molecular Cell Biology and Genetics, Dresden, 01307 Germany}
\author{Frank Jülicher}
\email{julicher@pks.mpg.de}
\affiliation{Max Planck Institute for the Physics of Complex Systems, Nöthnitzer Straße 38, 01187 Dresden, Germany}
\affiliation{Center for Systems Biology Dresden, Pfotenhauerstraße 108, 01307 Dresden, Germany}
\affiliation{Cluster of Excellence Physics of Life, TU Dresden, Dresden, 01307 Germany}
\author{Jan Brugués}
\email{jan.brugues@tu-dresden.de}
\affiliation{Cluster of Excellence Physics of Life, TU Dresden, Dresden, 01307 Germany}
\affiliation{Max Planck Institute for the Physics of Complex Systems, Nöthnitzer Straße 38, 01187 Dresden, Germany}
\affiliation{Max Planck Institute of Molecular Cell Biology and Genetics, Dresden, 01307 Germany}

\bibliographystyle{unsrt}  

\begin{abstract}

Oscillatory biochemical signals and mechanical forces must coordinate robustly during development, yet the principles governing their mutual coupling remain poorly understood. In syncytial embryos and cell-free extracts, mitotic waves propagate across millimeter scales while simultaneously generating cytoplasmic flows, suggesting a two-way interaction between chemical oscillators and mechanics. Here, we combine experiments in \textit{Xenopus laevis} cytoplasmic extracts with a minimal particle-based model to reveal a mechanochemical feedback that stabilizes phase wave propagation. In contrast to previous models of oscillatory active matter, an asymmetric size cycle—slow growth and rapid shrinkage—combined with size-dependent mechanical interactions generates a net particle displacement and flows aligned with the wave direction, which in turn drive a synchronization transition. Our results show that mechanical forces actively maintain the coherence of biochemical waves, providing a general mechanism for long-range order in oscillating active matter.

\end{abstract}

\maketitle

\section*{Introduction} 
Biological development requires the rapid and reliable emergence of  precise spatial and temporal coordination.
Spatial organization is achieved through compartmentalization and mechanical forces~\cite{heisenberg2013forces, martin2009pulsed, rinaldin2024robust}, while biochemical oscillators provide temporal control over processes such as cell division and gene expression~\cite{novak1993modeling, murray1989cyclin, rohde2024cell, ho2024nonreciprocal}. A key question in developmental biology and biophysics concerns how these physical principles interact robustly: mechanical forces generate flows that advect chemical signals~\cite{goldstein2015physical, kim2018coordination}, while oscillators modulate when and where forces are applied~\cite{deneke2019self}. Systems exhibiting such an interplay are examples of oscillating active matter~\cite{zhang2023pulsating,mackay2026active,dierkes2014spontaneous}. Understanding this interplay is essential for explaining how living systems achieve reliable organization despite intrinsic noise and external perturbations~\cite{perez2016intrinsic, romeo2025information}.

This interplay is particularly relevant during early embryogenesis in syncytial systems, where multiple nuclei share a common cytoplasm. In organisms such as Drosophila embryos and \textit{Xenopus laevis} egg extracts, spatial organization emerges without physical boundaries and is coordinated by traveling waves of Cdk1 activity that synchronize mitotic entry over millimeter scales~\cite{deneke2016waves, chang2013mitotic}.
These biochemical waves are accompanied by cytoplasmic flows and nuclear rearrangements~\cite{lv2020emergent, deneke2019self}, raising the possibility that cell cycle oscillators directly shape the system’s mechanical organization. Yet most studies treat the mitotic wave as an externally imposed signal~\cite{koke2014computational}, leaving open the question of whether mechanical feedback plays a role in stabilizing wave propagation itself.

So far, theoretical frameworks typically consider stationary nuclei, described either as sources of diffusible activators in reaction--diffusion systems~\cite{nolet2020nuclei, gelens2014spatial} or as Kuramoto-coupled phase oscillators~\cite{mishra2026geometry}. In syncytia, however, nuclei are advected by cytoplasmic flows~\cite{lv2020emergent}, exert mechanical forces on one another~\cite{hur2025topological,koke2014computational}, and might exchange neighbors over time. Mechanical activity and motility are known to strongly affect synchronization: competition for space can suppress phase coherence~\cite{zhang2023pulsating}, while motility can enhance synchronization by promoting mixing and neighbor exchange~\cite{rouzaire2025activity, uriu2013dynamics}. Conversely, even in the absence of explicit phase coupling, oscillatory size changes can fluidize dense assemblies and modify transport properties~\cite{tjhung2017discontinuous}. Capturing these competing effects requires models in which oscillations feed back on mechanics, and mechanics in turn transports and couples oscillators. 
A step in this direction is offered by a two-fluid model of the Drosophila embryo, where the cell cycle oscillator drives cortical actomyosin contractility and the resulting flows advect nuclei, whose position in turn reshapes the contractile field~\cite{hernandez2023two}; however, the oscillator's phase is imposed as a single global clock, so mechanics can relocate nuclei but cannot itself desynchronize, entrain, or otherwise act on their phase in this formulation.
Therefore a fundamental open question remains: how do chemical oscillators and mechanical forces jointly maintain robust long-range coordination in a noisy, advecting environment?

Here, we uncover a minimal mechanochemical mechanism that stabilizes wave propagation by coupling oscillatory phase dynamics to size-dependent mechanical interactions. Using \textit{Xenopus laevis} cytoplasmic extracts, a well-established system that recapitulates mitotic waves in a controlled setting~\cite{chang2013mitotic, murray1989cyclin, afanzar2020nucleus}, we quantify particle-scale flows and cell cycle phase, revealing that flows are directed along the direction of wave propagation. We then introduce a model in which particles grow slowly and shrink rapidly under cell cycle control. This asymmetric size cycle, combined with repulsive mechanical interactions, generates flows aligned with the wave, which in turn enhance particle mixing and phase synchronization. The resulting feedback produces a robust transition from a flowing, weakly synchronized regime to a synchronized state with strongly reduced flows. Our results show that mechanically generated flows do not destabilize the wave but instead stabilize it, providing a generic mechanism for long-range order in oscillating active matter.

\section{Cell cycle waves and cytoplasmic flows are coupled in cytoplasmic extracts}
\begin{figure*}
    \centering
    \includegraphics[width=\linewidth]{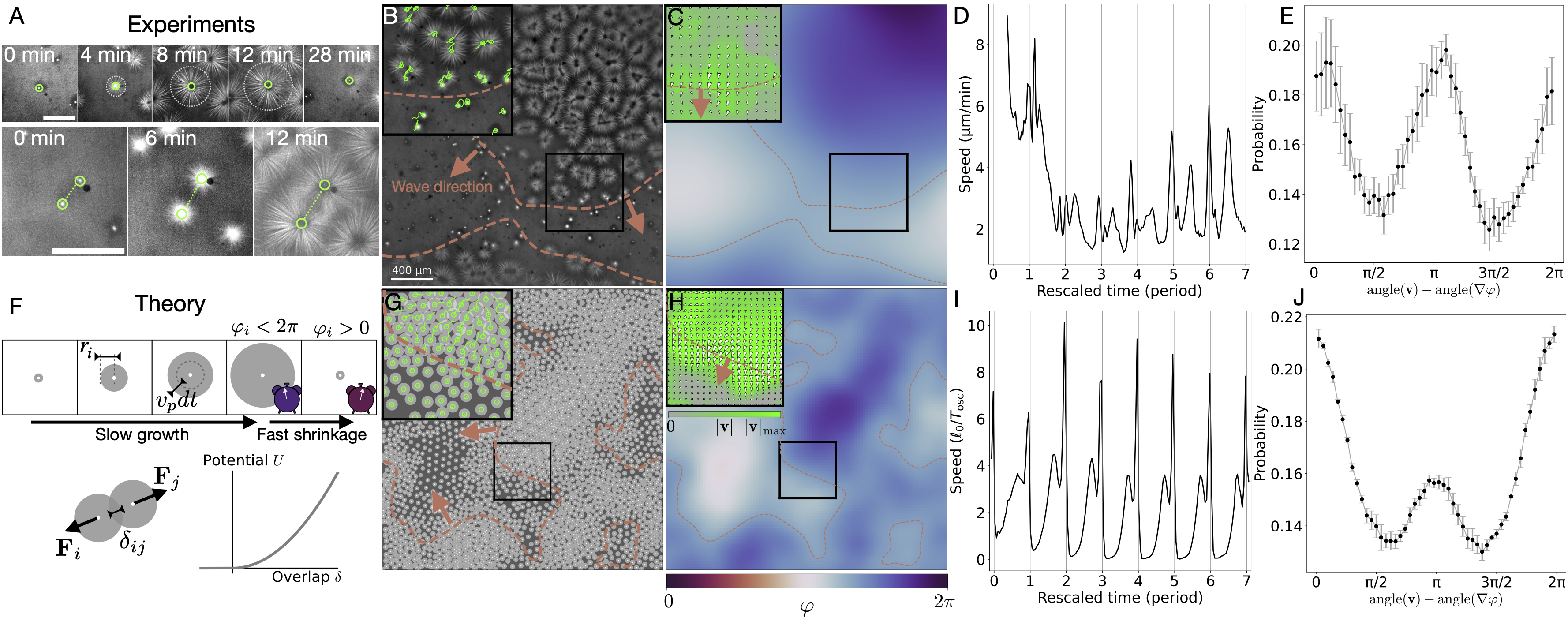}
    \caption{\textbf{Centrosome flows driven by microtubule growth dynamics are phase-locked to the mitotic wave in cytoplasmic extract.}
    (A)~Cropped microscopy images (labeled tubulin) showing asters that grow, shrink, and mechanically interact. Scale bars $200~\mu m$.
    (B)~Tubulin intensity snapshot of the extract, showing alternating polymerized (bright) and depolymerized (dark) regions. The dotted orange line marks an iso-phase contour of the mitotic wave; the arrow shows its propagation direction. Inset: bead trajectories (green) showing correlated flows.
    (C)~Experimental phase field extracted from microtubule intensity, with the corresponding bead-derived velocity field (inset). Dashed lines mark iso-phase contours.
    (D)~Mean bead speed as a function of local phase, expressed as rescaled time (fraction of the elapsed cell cycle).
    (E)~Probability density of the angle between the local phase gradient (opposite to the wave direction) and the local bead velocity, averaged per experiment and across 30 experiments from 8 extract preparations (1–8 cell cycles each). 
    (F)~Schematic of the agent-based model: two-dimensional particles grow at rate $v_p$ and
    depolymerize instantaneously when their internal phase $\varphi_i$ reaches $2\pi$, while interacting via a soft repulsive potential $U$ giving repulsive forces $\mathbf{F}_i=-\mathbf{F}_j$.
    (G)~Simulation snapshot of $N=2500$ particles exhibiting self-organized dense and dilute regions; inset: particle tracks. ( \Gr $=1.2$, $\omega=1$, $\sigma=0.2$, $\epsilon=0.5$, $D_\varphi/\epsilon=0.01$, $\mu U_0=2$, $r_0/\ell_0=0.05$)
    (H)~Simulated phase and velocity (inset) fields, coarse-grained from particle phases and velocities; the dashed pink line marks an iso-phase contour computed from the coarse-grained phase field.
    (I)~Mean particle speed as a function of local rescaled time, as in (D), for the parameters in (G).
    (J)~Probability density of the angle between the local phase gradient and local velocity, as in (E), averaged over all particles and three simulations with parameters as in (G).}
    
    \label{fig:xp_model}
\end{figure*}

To study the interplay between a chemical oscillator and mechanical flows, we use cell-free cycling cytoplasmic extracts from \textit{Xenopus laevis} eggs, supplemented with Aurora kinase A-coated (AurkA) beads, which act as artificial centrosomes~\cite{tsai2005aurora,field2019disassembly}.
Each bead leads to the nucleation of microtubules and forms the center of an aster~\cite{field2019disassembly,pelletierComovementAstralMicrotubules2020,field2018assembly}. 
As shown in Fig.~\ref{fig:xp_model}A, asters first grow slowly as microtubules polymerize and then shrink rapidely upon depolymerization. Two neighboring growing asters  can exert forces on each other that push them apart (Fig.~\ref{fig:xp_model}A). Polymerization cycles are governed by the cell cycle oscillator, which sets the timing of microtubule growth and shrinkage. As a consequence, each aster undergoes size oscillations and the cell cycle functions as an internal clock that synchronizes depolymerization. This synchronization leads to mitotic waves of polymerization and depolymerization that travel through the extract (Fig.~\ref{fig:xp_model}B, and Supplementary video S1). These waves appear as bright polymerized regions separated from dark depolymerized regions by a well-defined wave front (dotted line in Fig.~\ref{fig:xp_model}B). These oscillations are accompanied by aster flows, evidenced by the tracks of the beads (Fig.~\ref{fig:xp_model}B inset). Taken together, these features make the extract an ideal system to study the interplay of chemical and mechanical dynamics.

The system allows independent access to the chemical and mechanical degrees of freedom, with the oscillator phase inferred from the microtubule intensity field and the flows quantified from the trajectories of fluorescent AurkA beads at aster centers (Supplementary video S1). 
To quantify the local state of the cell cycle, we compute a phase field $\varphi(\mathbf{x},t)$ from this intensity signal (Fig.~\ref{fig:xp_model}C, SI section IIB and Fig.~S1): neighboring pixels are cross-correlated to obtain their relative timing, which is then globally combined and normalized between successive depolymerization events to yield a continuous phase $\varphi(\mathbf{x},t) \in [0,2\pi)$. The resulting field measures how advanced each region is in the chemical cycle, offering a robust way to quantify phase in systems where direct measurements of Cdk1 activity, and therefore standard phase-evaluation techniques, are unavailable~\cite{tan2020topological}.

To measure the flows, we track the motion of the AurkA beads to determine a local velocity field $\mathbf{v}(\mathbf{x},t)$ (Fig.~\ref{fig:xp_model}C inset). Bead positions are localized frame-by-frame and linked into trajectories (see SI Section IIA), from which the local velocity field is obtained by finite differencing bead displacement, providing a direct read-out of the cytoplasmic mechanical flow around each aster.
The extract exhibits spatially correlated flows that are strongly modulated by the cell cycle phase, with one or two peaks of the average bead speed observed per cycle (Fig.~\ref{fig:xp_model}D), indicating that the mechanical activity of the system is gated by the biochemical oscillator (validated in SI, Fig.~S2). We next ask whether the flow orientation is also coupled to the wave: computing the local phase gradient $\mathbf{\nabla}\varphi$ (SI, Fig.~S3) reveals that bead velocities are either parallel or antiparallel to $-\mathbf{\nabla}\varphi$, the wave direction (Fig.~\ref{fig:xp_model}E, Fig.~S4, Supplementary video S2).

\section{Minimal mechanochemical model reproduces the wave-flow coupling}

Experiments show that the cell cycle oscillator induces flows oriented with the wave direction. To explain these observations, we develop a two-dimensional (2D) agent-based model that captures the essential features of the system. 
We describe aster growth as a linear increase of the particle radius $r_i$ at a constant rate
\begin{equation}
\dot{r}_i = v_p,
\label{eq:radius_dynamic}
\end{equation}
and microtubule depolymerization as the reset of the particle size to $r_0$ once its internal phase $\varphi_i$ reaches $2\pi$ (Fig.~\ref{fig:xp_model}F). The phase advances at a rate $\omega$. Importantly, phase and growth control a different aspect of the particles' dynamics: the cell cycle oscillator controls depolymerization timing, while microtubule aster polymerization sets the growth rate. This distinction allows us to vary the relative timescales of chemical and mechanical processes. We quantify this timescale separation using a dimensionless control parameter, the \emph{growth number},
\begin{equation}
\frac{T_{\mathrm{osc}}}{T_{\mathrm{grow}}}
             = \frac{v_p T_{\mathrm{osc}}}{\ell_0},
\qquad 
\ell_0 = \frac{L}{\sqrt{N}},
\label{eq:Gr_definition}
\end{equation}
which compares the chemical oscillation period $T_{\mathrm{osc}}=2\pi/\omega$ to the time 
$T_{\mathrm{grow}}$ needed for a particle to grow by one interparticle spacing 
$\ell_0$. When \Gr~$\ll 1$, particles remain dilute throughout the cycle 
and flows are negligible because they do not interact mechanically; when  \Gr~$\gtrsim 1$, particles are in contact with each other at each cycle, generating flows. The growth number therefore quantifies how much a particle can grow during one chemical cycle relative to the typical distance separating neighboring particles. By definition, the growth number is also the ratio of length scales $(r_\mathrm{max}-r_0)/\ell_0$, where $r_\mathrm{max}$ is the maximal particle radius reached at the end of a cycle. 
In the limit $\ell_0\gg r_0$, which is the regime explored in our simulations, \Gr$\sim r_\mathrm{max}/\ell_0$. This parameter also characterizes the extent of particle overlap at the end of the cycle.
Throughout the manuscript, lengths are expressed in units of the typical interparticle spacing, $\ell_0 $, which corresponds to the mean distance between particles in a 2D domain of side length $L$ containing $N$ particles. Because particle sizes vary over the oscillation cycle, $\ell_0$ provides a fixed and physically meaningful length scale that does not depend on instantaneous particle radii. Time is expressed in units of the chemical period $T_{\mathrm{osc}}$ so that one oscillation cycle corresponds to $t=1$. We represent mechanical interactions through a soft repulsive potential $U=\frac{1}{2}\,U_0\,\delta^2$ between particles that overlap by a distance $\delta$ (Fig.~\ref{fig:xp_model}F). A repulsive force $\mathbf{F}_{ij}$ derives from this potential and is nonzero when particle $i$ and particle $j$ overlap. The resulting overdamped dynamics of particle positions $\mathbf{x}_i$ are given by
\begin{equation}
\dot{\mathbf{x}}_i =
    \frac{\mu}{r_i}\,\sum_{j\neq i}\mathbf{F}_{ij},
    \qquad
    \mathbf{F}_{ij} = 
    -U_0\,\delta_{ij}\,
    \mathcal{H}(\delta_{ij})\,\mathbf{e}_{ij},
\label{eq:pos_dynamic}
\end{equation}
where $\delta_{ij}=r_i+r_j-||\mathbf{x}_i-\mathbf{x}_j||$ is the overlap distance, $\mathbf{e}_{ij}$ is the unit vector pointing from particle $i$ to particle $j$, and $\mathcal{H}$ is the Heaviside function. The size dependence of aster motility is captured through a mobility $\mu/r_i$, motivated by the Stokes’ law scaling of viscous drag.

Finally, to propagate the biochemical wave through the discrete system, the particle phases, defined modulo $2\pi$, obey a noisy Kuramoto coupling~\cite{acebron2005kuramoto}:
\begin{equation}
\dot{\varphi}_i = \omega_i + \epsilon \sum_{\langle i,j\rangle} \sin(\varphi_j - \varphi_i)
    + \sqrt{2D_\varphi}\,\eta_i ,
\label{eq:kuramoto}
\end{equation}
where $\langle i,j\rangle$ denotes the neighbors of particle $i$, identified via a Delaunay construction \cite{VandenHeuvel2024DelaunayTriangulation}, and $\eta_i$ is the Gaussian white noise. This coupling represents the chemical interaction between neighboring asters, corresponding to the spatial homogenization of Cdk1 activity. The intrinsic frequencies $\omega_i$ are drawn from a normal distribution with mean $\omega$ and standard deviation $\sigma$. Because the phase cycle governs physical properties of the particles (in particular their size), $\omega$ cannot be eliminated by transforming to a corotating frame, as is commonly done for Kuramoto oscillators. A related model was introduced in Ref.~\cite{zhang2023pulsating}. In our model, the Kuramoto coupling arises from the diffusion of a biochemical oscillator and is directly motivated by experimental observations. In addition, our model incorporates a size-dependent motility and an asymmetric size cycle, which introduce qualitatively new dynamical effects discussed below.

We perform numerical simulations of $N$ particles initialized with random positions and random phases. After a few oscillation periods, the system self-organizes into propagating waves of polymerization and depolymerization, accompanied by the emergence of dense and dilute regions (Fig.~\ref{fig:xp_model}G), reflecting particle size and phase  (Fig.~\ref{fig:xp_model}H). Particle's speed is maximal at the end of the cell cycle (Fig.~\ref{fig:xp_model}I), and the resulting flow field is strongly aligned with the direction of wave propagation (Fig.~\ref{fig:xp_model}J, Supplementary video S3).
Despite this overall agreement, we note one quantitative difference between simulations and experiments. In the extract, residual flows persist even in fully depolymerized regions, whereas our dry agent-based model predicts negligible motion at near zero phases. In particular, Fig.~\ref{fig:xp_model}D shows non-zero speeds at the beginning of the cycle in the experiment, a feature absent from the simulations. These residual flows are likely mediated by hydrodynamic coupling within the cytoplasm, which is not included in our minimal model. Overall, the simulations capture the essential qualitative features of the experiments while isolating the minimal ingredients required for phase-locked flows.

\section{Phase gradient leads to flow}
\label{sec:phase_gradient}

\begin{figure*}
    \centering
    \includegraphics[width=\linewidth]{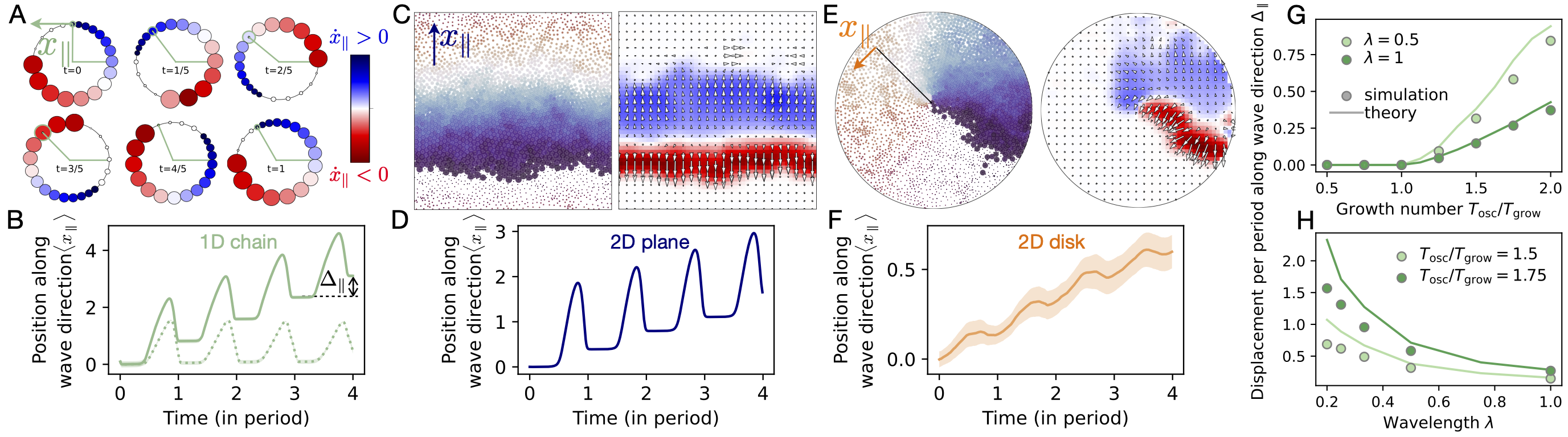}
    \caption{\textbf{Particles flow along the phase gradient.}
    (A)~One-dimensional chain with periodic boundaries. A phase wave of wavelength $L$ induces periodic growth and shrinkage of the particles. When particles are in contact, they form a dense cluster whose front moves in the direction of the wave and whose rear moves in the opposite direction. Colors indicate instantaneous velocity relative to the wave direction (blue: trigonometric; red: opposite). The circled particle shows a net forward drift over one cycle.
    (B)~Mean particles' positions, $\langle x_\parallel(t)\rangle $ (Eq.~\ref{eq:x_parallel_definition}) averaged in the comoving frame. Particles exhibit oscillatory motion with a net drift $\Delta_\parallel$ per cycle. The dashed green curve shows the 1D system without size-dependent mobility.
    (C)~2D system with periodic boundaries. A planar wave is imposed along one box direction. Left: particle phases (same color map as Fig.~\ref{fig:xp_model}); right: arrows show the particle velocity, and the color map show the values of velocity projected along the wave direction. A dense band forms where particles are large and in contact.
    (D)~Mean particle's positions projected along the wave direction (the y-axis) averaged in the comoving frame. 
    (E)~2D circular system with a central phase defect. The phase winds around the defect, generating a tangential phase gradient. A dense wedge forms and particles flow along this tangential gradient.
    (F)~Mean particle's positions projected along the wave direction (the tangential direction averaged in the comoving frame. 
    (G--H)~Net drift $\Delta_\parallel$ (Eq.~\ref{eq:delta_parallel_definition}) as a function of the growth number \Gr~or the wavelength $\lambda$ calculated from the semi-analytical theory (SI Section IV) (solid lines) and from 1D simulations (circles). Drift occurs only when particles grow sufficiently during one oscillation period to come into contact with their neighbors, and decreases with the wavelength. }
    \label{fig:grad_to_flow}
\end{figure*}

Because phase correlates with particle size and therefore with local packing, this suggests a simple physical principle: particles tend to flow from densely packed (late-phase) regions toward more dilute (early-phase) regions. Here, we test and quantify this hypothesis by analyzing minimal geometries in which the phase profile is externally imposed or strongly constrained. These controlled settings allow us to isolate the mechanism by which a phase gradient generates directed particle motion.

\paragraph*{One-dimensional (1D) chain.} We first consider a 1D chain simulation of particles with periodic boundaries. We initialize the system with a linear phase profile $\varphi(x)=2\pi x/L$, such that the phase increases uniformly over the domain of length $L$. For sufficiently strong coupling $\epsilon$ in Eq.~\ref{eq:kuramoto}, this profile remains phase-locked and produces a traveling wave of period $T = 2\pi/\omega$ and wavelength $\lambda = L$. In this configuration, particle sizes vary smoothly along the chain, and particles periodically switch between a \emph{dilute} state, where they do not touch their neighbors, and a \emph{dense} state, where they form a compact cluster (Fig.~\ref{fig:grad_to_flow}A). In the dilute state, particles do not experience interaction forces and essentially do not move. In the dense state, they are in contact, and repulsive forces generate a collective motion within the cluster. This dense cluster grows in size along the wave direction because of particle growth, resulting in its front advancing and its rear receding: particles located near the front move forward, whereas those near the back move backward. An individual particle thus undergoes an oscillatory trajectory over successive cycles: when the dense region reaches it, it is advected forward or backward depending on its position within the region, and once depolymerization occurs the system becomes dilute and particle motion ceases. We quantify this displacement by averaging the particle position in the comoving frame 
\begin{equation}
\langle x_\parallel \rangle = \frac{1}{N}\sum_i x^{(i)}_\parallel (t-\frac{T}{\lambda} x^{(i)}_\parallel(t=0))
\label{eq:x_parallel_definition}
\end{equation}
We show $\langle x_\parallel \rangle$ calculated from the 1D chain simulation with $N=50$ particles in Fig.~\ref{fig:grad_to_flow}B and visualize how the particles positions oscillates in time. These forward and backwards displacements result in the two velocity peaks of Fig.~\ref{fig:xp_model}I.

Because mobility depends on size (Eq.~\ref{eq:pos_dynamic}), the forward and backward displacements experienced during the dense part of the cycle do not cancel exactly. This asymmetry in the cycle effectively breaks time-reversal symmetry, in analogy with the scallop theorem~\cite{purcell2014life}. As a result, each particle acquires a net drift. We quantify this net drift as the averaged displacement projected onto the phase-gradient direction over one oscillation period,
\begin{equation}
\Delta_{\parallel} \;=\; \langle x_{\parallel}  (t+T_\mathrm{osc} \rangle- \langle x_{\parallel}(t) \rangle,    
\label{eq:delta_parallel_definition}
\end{equation}
As shown in Fig.~\ref{fig:grad_to_flow}B (green curve), $\Delta_\parallel$ is positive in the 1D system, indicating a net drift along the phase gradient. Size-dependent mobility is essential for generating this net drift. If we replace Eq.~\ref{eq:pos_dynamic} by $\dot{x}_i = -\mu F_i$ (constant mobility), particles still form dense clusters and undergo oscillatory motion while in contact, but the forward and backward contributions within each cycle cancel (dashed green curve in Fig.~\ref{fig:grad_to_flow}B). Thus, directed flow arises from the combination of (i) periodic formation and dissolution of dense clusters due to growth and depolymerization, and (ii) size-dependent mobility within these clusters.

\paragraph*{2D systems.} We next test whether this mechanism persists in two dimensions. In Fig.~\ref{fig:grad_to_flow}C, a planar wave is imposed along one box direction in a periodic domain. As in the 1D case, particles periodically grow and shrink in response to the phase profile and alternately form dilute (particle colored in orange and beige) and dense (particles colored in purple and blue) regions. When particles are large and in contact, they organize into a dense band aligned with the phase front. Within this band, particles experience repulsive forces and flow along the phase gradient; outside the band, in the dilute regions, they barely move, as shown from the flow field of Fig.~\ref{fig:grad_to_flow}C. The resulting velocity field has a nonzero projection along the phase gradient, and the displacement $x_\parallel(t)$ again displays oscillations with a nonzero net drift per cycle (Fig.~\ref{fig:grad_to_flow}D). 
The same behavior occurs in the presence of a phase topological defect and particles confined in a circular disk with soft boundaries. In Fig.~\ref{fig:grad_to_flow}E, a central phase singularity creates a tangential phase gradient. Particles now assemble into a dense wedge when they are large, and flows are localized in this wedge. The induced motion is predominantly tangential, and the tangential displacement exhibits a cumulative drift over time (Fig.~\ref{fig:grad_to_flow}F). This demonstrates that the coupling between flow and phase gradient is geometric and does not depend on planar wave geometry or boundary conditions. 

\paragraph*{Dependence on growth number and wavelength.}
We now quantify how the net drift depends on the amount of growth occurring during one oscillation period. The growth number \Gr~(Eq.~\ref{eq:Gr_definition}) sets the magnitude of mechanical interactions between particles. When \Gr~is smaller than one, particles do not grow sufficiently within one period to come into contact before depolymerization, and particles experience negligible motion, leading to $\Delta_\parallel \approx 0$ (circles in Fig.~\ref{fig:grad_to_flow}G). As \Gr~increases further, particles spend more time in contact and experience stronger mechanical interactions, which enhances $\Delta_\parallel$.  
We also vary independently the wavelength $\lambda$ of the phase profile while keeping the oscillation period fixed. We find that the net drift $\Delta_\parallel$ decreases as $\lambda$ increases, even though the local yo-yo motion persists (circles in Fig.~\ref{fig:grad_to_flow}H). This behavior can be understood from the spatial structure of the dense region. For shorter wavelengths, the phase gradient is steeper, leading to a more densely packed region in which particles experience stronger and more asymmetric mechanical interactions over a cycle. As $\lambda$ increases, the dense region spreads out, and the asymmetry between forward and backward displacements is reduced, resulting in a smaller net drift. We confirm these results by numerically solving the reduced 1D dynamics of the growth-induced mechanical interactions (solid lines in Fig.~\ref{fig:grad_to_flow}G-H, see SI Section IV and Fig. S5). This reduced description considers a one-dimensional chain of growing, overlapping particles and tracks the overlap profile of a bundle of $N_c$ mutually contacting particles as they grow, are pushed apart, and are periodically replaced by depolymerization at one edge and growth at the other. Because growth continuously injects overlap while elastic repulsion relaxes it only at the free edges of the bundle, the resulting overlap profile — and hence the particle velocity profile — is inherently asymmetric, producing a net drift per cycle that increases with the growth number and decreases with $\lambda$, consistent with the full 2D simulations. In summary, the net particle drift over one cycle is set by growth-induced mechanical interactions and controlled jointly by the growth number and the spatial properties of the propagating wave.

\section{Mechanically controlled synchronization transition}

\begin{figure*}
    \centering
    \includegraphics[width=\linewidth]{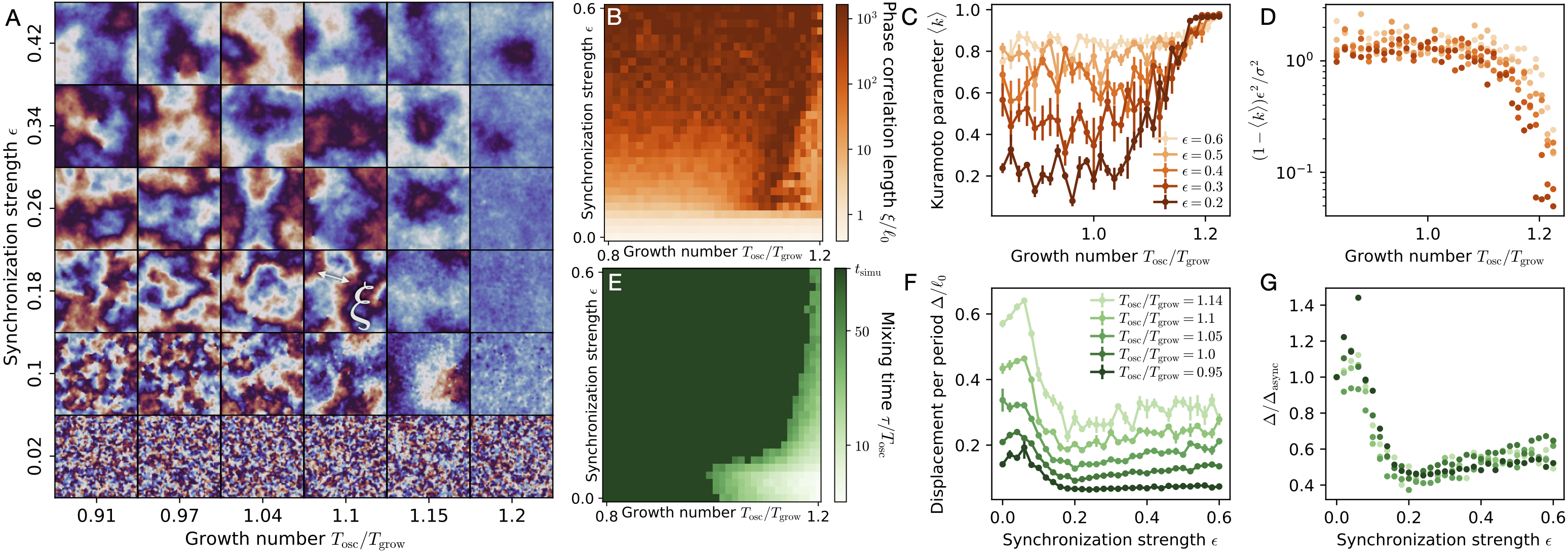}
    \caption{\textbf{Mechanical–oscillatory coupling maintains order in oscillating active matter.}
    (A)~Steady-state coarse grained phase fields (colorscale as in Fig. \ref{fig:xp_model}) for varying synchronization strength $\epsilon$ and growth number $T_\mathrm{osc}/T_\mathrm{grow}$. The other parameters are indicated in Fig.~\ref{fig:xp_model}. The images are shown after $120$ periods ($t_\mathrm{simu}=120\,T_\mathrm{osc}$).
    (B)~Phase diagram of the correlation length of the phase field $\xi$. Spatial coherence increases with synchronization until full phase alignment is reached, after which $\xi$ decreases again.
    (C)~Kuramoto parameter $\langle k \rangle$ (Eq.~\ref{eq:kop_definition}) as a function of the growth number, revealing a mechanically induced synchronization transition.
    (D)~Rescaling the Kuramoto parameters with the synchronization strength collapses the curve for values of $\epsilon$ larger than $\sigma=0.2$ (colors as in (C)).
    (E)~Phase diagram of the neighbor-exchange timescale $\tau$ (color, logscale). At high \Gr, particles exchange neighbors rapidly, indicating substantial mixing.
    (F)~The average displacement after one period $\Delta/\ell_0$ (Eq.~\ref{eq:delta_definition}) with synchronization strength, indicating that synchronization suppresses diffusive motion. This is consistent with observations of the previous section: increasing $\epsilon$ increases the typical length scale, and this decreases the net displacement.
    (G)~Rescaling $\Delta$ by its value in the asynchronzied regime reveals a chemically induced slow down transition. (colors as in (F).}
    \label{fig:synchro_transition}
\end{figure*}

We have shown that a phase gradient and an asymmetric cycle generate directed particle motion when the growth number is large enough for particles to transiently come into mechanical contact. We now generalize this analysis to systems initialized with fully disordered phase profiles. In this regime, the interplay between mechanically induced flows and phase coupling leads to rich self-organization: depending on parameters, the system either develops finite-sized synchronized domains or reaches global phase synchronization.

\paragraph*{Phase coherence in the presence of flow.}
We investigate how the steady-state spatial coherence of the phase field depends on the synchronization strength $\epsilon$ and the growth number \Gr. Increasing \Gr~enhances mechanically induced particle motion, which can in principle either disrupt phase coherence through mixing or promote synchronization by coupling distant regions of the system.
Representative steady-state phase fields are shown in Fig.~\ref{fig:synchro_transition}A, revealing three distinct regimes. For low synchronization strength $\epsilon$, the system remains largely asynchronous. At larger $\epsilon$ and small \Gr, the system develops synchronized domains of finite extent, characterized by a well-defined correlation length smaller than the system size. Finally, for sufficiently large $\epsilon$ and large \Gr, the system becomes fully synchronized.
To quantify this behavior, we compute the spatial correlation function of the coarse-grained phase field and extract the correlation length $\xi$ by fitting its decay to an exponential. Lengths are expressed in units of the typical interparticle spacing $\ell_0$ (Fig.~\ref{fig:synchro_transition}B, SI Section IIIB). 
As \Gr~increases, $\xi$ initially grows and eventually exceeds the system size, indicating long-range coherence. For \Gr~and $\epsilon$ very large, however, $\xi$ decreases again. In this regime the system is fully synchronized, spatial phase gradients vanish, and $\xi$ ceases to represent a meaningful wave length scale. This non-monotonic behavior is consistent with the presence of a synchronization transition.
To directly characterize synchronization, we compute the Kuramoto parameter at steady state
\begin{equation}
k  = \frac{1}{N} \left| \sum_{j=1}^{N} e^{i \varphi_j} \right|,
\label{eq:kop_definition}
\end{equation}
and its average $\langle k \rangle$ over four simulations and $3$ periods at steady state. As shown in Fig.~\ref{fig:synchro_transition}C, $\langle k \rangle$ increases with both the intrinsic phase-coupling strength $\epsilon$ and the growth number \Gr. 
Remarkably, rescaling the level of asynchrony $1-\langle k \rangle$ by the square of the intrinsic coupling strength $\epsilon^2$ leads to a collapse of the data obtained at different $\epsilon$ (Fig.~\ref{fig:synchro_transition}D, SI Section IIID). This $\epsilon^2$ dependence is also found in a mean-field Kuramoto model, where deep in the phase-locked regime, the typical phase spread scales as $\sigma/\epsilon$, giving $1-\langle k \rangle \sim \sigma^2/\epsilon^2$. This type of scaling can also hold beyond mean-field coupling~\cite{hong2005collective} (SI Section IIID). This scaling indicates that $\epsilon$ primarily sets the overall level of synchrony, and thereby the correlation length, while the onset of synchronization is controlled by the mechanically induced mixing encoded in the growth number \Gr~in this regime. Such data collapse is consistent with the presence of a robust synchronization transition governed by a single dominant control parameter. Using the measured aster growth rate, oscillation period, and mean interparticle spacing in cytoplasmic extracts (Table S1, SI Section IIE), we estimate a growth number \Gr~$\approx 4$ for our experimental conditions, well above the \Gr~$\geq1$ threshold associated with the synchronized regime in the model. While this comparison is only qualitative, given the simplified nature of our model, it is consistent with the fact that mitotic waves in extracts propagate coherently over large distances, and suggests that the extract naturally operates in a regime where mechanical feedback would be expected to support synchronization.

\paragraph*{Neighbor exchange and the onset of an effective many-body coupling.}
To uncover the origin of this mechanically assisted synchronization, we quantify how frequently particles exchange neighbors, which provides a measure of particle mixing. At each time point, we construct the adjacency matrix $M_{ij}$ associated with the Delaunay triangulation and evaluate its temporal autocorrelation and extract the typical mixing time $\tau$ by fitting the autocorrelation decay to an exponential (SI section IIIC).
Figure~\ref{fig:synchro_transition}D shows $\tau$ across parameter space. Systems beyond the synchronization transition, corresponding to large \Gr~and intermediate or large $\epsilon$, exhibit rapid neighbor exchange, with $\tau/T_\mathrm{osc}$ shorter than the simulation duration. In this regime, particle motion effectively introduces an effective many-body coupling: because neighbors continuously reshuffle, each particle interacts with many others over a few oscillation periods. This enhanced mixing provides a natural explanation for the mechanically induced synchronization observed at large \Gr.

Finally, we quantify the magnitude of mechanically induced motion by measuring the mean particle displacement after one oscillation period, 
\begin{equation}
    \Delta/\ell_0 = \sqrt{\langle r^2(t=T_{\mathrm{osc}})\rangle}/\ell_0.
\label{eq:delta_definition}
\end{equation}
As expected, increasing \Gr~leads to larger displacements, reflecting stronger mechanically induced flows (Fig.~\ref{fig:synchro_transition}E). Interestingly, the synchronization strength $\epsilon$ also affects the flow magnitude: $\Delta$ decreases as $\epsilon$ increases, despite $\epsilon$ controlling only phase coupling (Fig.~\ref{fig:synchro_transition}F).
This reduction of flow with increasing $\epsilon$ arises because synchronization suppresses short-wavelength phase gradients, which are the source of mechanical motion. As shown earlier, particle flows are strongest when phase gradients are steep and occur over short length scales; increasing $\epsilon$ increases the spatial coherence length $\xi$, thereby reducing local gradients and weakening flows. Consistent with this interpretation, rescaling $\Delta$ by its value at $\epsilon=0$ collapses the curves for different \Gr  (Fig.~\ref{fig:synchro_transition}G). Together, these results demonstrate a genuine mechanochemical feedback: asynchrony generates flows, flows promote synchronization, and synchronization in turn suppresses the flows that created it.

\section*{Discussion}

\paragraph*{} Our results reveal a mechanochemical feedback loop by which an oscillatory system coupled to mechanics self-organizes robustly. Phase gradients generate directed particle flows; these flows promote particle mixing and neighbor exchange, which enhances phase synchronization. Synchronization, in turn, suppresses phase gradients and therefore suppresses the flows that generated it. This feedback loop was revealed by a combination of experiment and theory: extracts provided direct evidence for phase-flow coupling, and theoretical analysis allowed us to identify the resulting synchronization transition, showing that the same local phase–flow coupling measured experimentally is sufficient to drive collective synchronization.  This closed feedback loop provides an intrinsic stabilizing mechanism: asynchrony generates motion, motion promotes synchrony, and synchrony eliminates motion. As a consequence, the system converges either to a flowing, weakly synchronized regime characterized by short spatial wavelengths, or to a globally synchronized regime with long-range phase coherence and negligible mechanical motion. This self-limiting behavior ensures robust coordination without fine-tuning, despite noise, advection, and mechanical activity.

\paragraph*{}The mechanically induced synchronization transition we report is reminiscent of previous studies in which motility or active transport enhances phase coherence by promoting mixing~\cite{uriu2013dynamics, rouzaire2025activity}. A key distinction, however, is that in those models mixing is introduced as an external control parameter, for instance through self-propulsion or imposed mobility. In contrast, in our system, mixing emerges endogenously from oscillatory size dynamics controlled by a biochemical clock. Mechanical motion is therefore not imposed, but generated by the chemical cycle itself. This places our system in a distinct class of oscillatory or pulsating active matter, where internal clocks directly generate mechanical activity~\cite{zhang2023pulsating}. This work extends synchronized oscillatory active matter by treating particle size and phase as independent dynamical variables, whose coupling is controlled by the growth number \Gr. This provides a novel mechanism for flow generation that relies on the asymmetry of the particle size cycle. While collective swimmers based on asymmetric cyclic deformations have previously been introduced as hypothetical minimal models~\cite{najafi2004simple}, and realized in engineered particle robots with externally imposed phase offsets~\cite{li2019particle}, here propulsion emerges collectively from the intrinsic physical properties and biochemical dynamics of the active particles themselves.
More broadly, our results connect to the general principles of mechanochemical pattern formation, where feedback between biochemical signals and mechanical forces can generate spatial and temporal structure across scales~\cite{rombouts2023forceful}.

\paragraph*{}From a biological perspective, this mechanism offers a plausible explanation for the robustness of mitotic waves in large, advecting cytoplasmic environments such as syncytial embryos and cell-free extracts. In these systems, nuclei and centrosomes are not stationary but are continuously displaced by cytoplasmic flows and mechanical interactions~\cite{deneke2016waves, lv2020emergent}. Our results suggest that such flows need not undermine synchronization; instead, they can actively contribute to it by enhancing effective coupling between distant oscillators. Once synchronization is achieved, the suppression of phase gradients naturally limits further mechanical activity, stabilizing the coordinated state. Our model is intentionally minimal and has several limitations. First, it neglects hydrodynamic interactions, treating the cytoplasm as a dry medium. This likely explains the residual flows observed experimentally in depolymerized regions, which are absent from the simulations. Incorporating long-range hydrodynamic coupling may extend the spatial range of flows~\cite{shamipour2021cytoplasm, deneke2019self, goldstein2015green} without altering the local mechanochemical feedback identified here. Second, the model does not include active contractility, which are known to play important roles in embryos~\cite{field2011bulk,telley2012aster}. Exploring how this mechanism interacts with hydrodynamics, contractility, and deformability will be an important direction for future work. 
Finally, our description treats centrosomes as instantaneously synchronizing pacemakers. This is a simplification, which is expected to be valid in the dense cytoplasmic environment where molecular diffusion and coupling are fast compared to the cell cycle period. A more detailed approach would explicitly model the reaction–diffusion dynamics of Cdk1 and associated regulators, as was done for stationary oscillators in~\cite{haerter2023synchronization}. For rapidly diffusing molecules, this explicit modeling only slightly reduces the synchronization order parameter, and we do not expect it to qualitatively alter the collective dynamics reported here. 
Geometric perturbations alone also have been shown to reorganize collective wave behavior in this same extract system, nucleating spiral waves that outcompete target patterns, without any mechanical flow or active transport involved~\cite{cebrian2024spiral}. Whether such diffusively propagating transitions interact with the mechanically driven feedback identified here is an open question for future work. Together, these observations suggest that the extract system supports multiple, non-mutually-exclusive routes — geometric, biochemical, and mechanical — to reshaping collective oscillatory dynamics, and that our minimal description captures one such route while remaining tractable enough to isolate its contribution.

\subsection*{Methods}
\paragraph*{}All Methods are described in detail in the Supporting Information. Briefly,

\paragraph*{Husbandry of experimental animals:}
Female \textit{Xenopus laevis} frogs were cared for and handled following established guidelines~\cite{reed2005guidance}.

\paragraph*{Preparation of Cytoplasmic Extracts and Buffers:}Cytoplasmic  extracts are prepared from \textit{Xenopus laevis} eggs following standard protocols~\cite{field2017xenopus,chang2018robustly}. Samples of extract were prepared on ice. The extract was supplemented with AurkA beads to serve as artificial centrosomes.

\paragraph*{Imaging of cytoplasmic extract samples:}Imaging was performed using a spinning disk confocal microscope (Olympus IX83 equipped with a CSU-W1 Yokogawa disk) connected to two Hamamatsu ORCA-Fusion BT Digital CMOS cameras (SD1).

\paragraph*{Artificial centrosome (AurkA beads) tracking:}Particle tracking was performed using a combination of TrackMate v7 \cite{ershov2022trackmate}, within FIJI/ImageJ 2.7.0 \cite{schindelin2012fiji}, and custom Python scripts.

\paragraph*{Phase reconstruction from microtubule intensity.} We extracted a local oscillator phase from the microtubule intensity field by using the timing of depolymerization events as phase markers: pairwise time delays between neighboring pixels were obtained from cross-correlation of their intensity traces, then combined via a global least-squares fit to build spatially resolved maps of successive depolymerization times. The phase at each position and time was defined as the fractional time elapsed between consecutive depolymerization events, yielding a continuous phase field $\varphi(\mathbf{x},t) \in [0, 2\pi)$.

\paragraph*{Simulations.} We numerically integrated the coupled mechanochemical model Eqs.~\ref{eq:radius_dynamic},\ref{eq:pos_dynamic},\ref{eq:kuramoto} for particles interacting mechanically on a two-dimensional domain with periodic boundary conditions, using overdamped dynamics and discrete depolymerization events triggered when each particle's phase reached $2\pi$. Short-range interactions were computed efficiently using cell lists and Verlet neighbor lists, and particle phases were coarse-grained onto a spatial grid with a Gaussian kernel to visualize phase fields and compute correlation functions.

\subsection*{Data sharing plans}

Code for experimental data analysis \url{https://github.com/lara-koehler/WaveFlowExperimentalAnalysis}

Data and simulation code will be made available after the review process.

\subsection*{Author Contributions}
E.S., L.K. and J.B. planned the experiments. E.S. performed the experiments. L.K. and E.S. analyzed the experimental data. L.K. designed the phase analysis method of the experimental data. L.K. designed and performed the theory and numerical simulations. L.K., E.S., F.J. and J.B. designed the project and wrote the manuscript.

\subsection*{Acknowledgements}
We thank Ylann Rouzaire for critical reading of the manuscript, and acknowledge insightful discussions with Maria Kharlamova, Adam Lamson, John D. Treado and Matteo Ciarchi. We thank Tim Mitchison and his group for kindly providing the anti Aurora Kinase A (AurkA) antibody and Urša Uršic for coating the AurkA beads. L.K. was supported by the MPI-PKS visitors program and a MSCA Postdoctoral fellowship. L.K., E.S., F.J. and J.B. acknowledge support from the Deutsche Forschungsgemeinschaft (DFG; German Research Foundation) under Germany’s Excellence Strategy — EXC-2068-390729961 — Cluster of Excellence Physics of Life of TU Dresden. 

\bibliography{biblio}

\clearpage
\setcounter{section}{0}
\setcounter{figure}{0}
\setcounter{table}{0}
\renewcommand{\thefigure}{S\arabic{figure}}
\renewcommand{\thetable}{S\arabic{table}}

\onecolumngrid

\section*{Supplementary information for Flow–wave coupling synchronizes oscillations in growing active matter}

\section{Experimental methods}

The experiments were conducted by preparing cytoplasmic egg extract of \textit{Xenopus laevis}, reconstituting microtubule asters within it, and imaging their dynamics. We first describe the husbandry and ethical approval of the experimental animals (Sec.~\ref{subsec:frogs}). We then detail the preparation of the buffers and cytoplasmic extracts used throughout the experiments (Sec.~\ref{subsec:extract_prep}). Next, we describe the preparation of artificial centrosomes used to nucleate microtubule asters in the extract (Sec.~\ref{subsec:aurka-beads}). Finally, we present the sample assembly and imaging conditions used to acquire the data (Sec.~\ref{subsec:sample_prep_imaging}).

\subsection{Husbandry of experimental animals}
\label{subsec:frogs}
Female \textit{Xenopus laevis} frogs were cared for and handled following established guidelines~\cite{reed2005guidance}. The frogs were sourced from Nasco (LM00531) and Xenopus1 (4800). All experimental procedures received approval from the local animal ethics committee (Landesdirektion Sachsen; license no. DD24-5131/367/9, 25-5131/521/12 and 25-5131/564/25) and adhered to the European Communities Council Directive 2010/63/EU, which governs the protection of animals used in scientific research, as well as the German Animal Welfare Act. 

\subsection{Preparation of Cytoplasmic Extracts and Buffers}
\label{subsec:extract_prep}

Cytoplasmic extracts from frog eggs were prepared using previously established protocols~\cite{chang2018robustly, field2018assembly}, with minor adjustments to meet the specific requirements of our experiments. To ensure consistency and reproducibility, several buffers were prepared in advance. The 10x Marc’s Modified Ringer’s (MMR) solution consisted of 1 M NaCl, 20 mM KCl, 10 mM MgCl\textsubscript{2}, 20 mM CaCl\textsubscript{2}, 1 mM EDTA, and 50 mM HEPES in MilliQ water. After adjusting the pH to 7.8 with NaOH, the solution was filter-sterilized and stored at room temperature. For the 20x \textit{Xenopus} Buffer (XB), 2 M KCl, 20 mM MgCl\textsubscript{2}, and 2 mM CaCl\textsubscript{2} were dissolved in MilliQ water, and the pH was adjusted to 7.7 using KOH before storing at 4~\textdegree{}C. A 1 M HEPES solution was also prepared in MilliQ water, with the pH adjusted to 7.7 after filter-sterilization, and stored at 4~\textdegree{}C. Additionally, a 2 M sucrose solution was filter-sterilized and stored at 4~\textdegree{}C.

Adult female frogs were primed for egg collection by administering 50 IU of pregnant mare serum gonadotropin (PMSG; 779-675, Covetrus) 3--8 days prior to the experiment. This was followed by an injection of 500 IU of human chorionic gonadotropin (hCG; lCG10-10VL, Sigma) one day before egg collection. Post-injection, the frogs were kept in 1x MMR solution at 16~\textdegree{}C for 18 hours. On the day of the experiment, MilliQ water (3 liters) was equilibrated to 16~\textdegree{}C for preparing buffers.

Several buffers were freshly prepared on the experiment day. The Dejelly Buffer was prepared by dissolving 20 g of L-cysteine (Merck) in 50 ml of 20x \textit{Xenopus} Buffer, and MilliQ water was added to a final volume of 1 liter. The pH was adjusted to 7.8 with NaOH. A 0.2x MMR solution was prepared by diluting 10x MMR in MilliQ water to a final volume of 1 liter, followed by pH adjustment to 7.8 using NaOH. The 1x \textit{Xenopus} Buffer was made by mixing 50 ml of 20x \textit{Xenopus} Buffer with 50 mM sucrose and 10 mM HEPES in MilliQ water to a final volume of 1 liter, with the pH adjusted to 7.7 using KOH. A calcium ionophore solution (CaIo) was prepared by adding 5~\textmu{}l of a 10 mg/ml calcium ionophore solution (Sigma) to 100 ml of 0.2x MMR buffer. Additionally, \textit{Xenopus} Buffer+ was prepared by adding 100~\textmu{}l of a 10 mg/ml solution of leupeptin (Thermo Scientific), pepstatin (Roth), and chymostatin (Calbiochem) to 100 ml of 1x \textit{Xenopus} Buffer.

\subsection{Artificial centrosomes preparation (AurkA beads)}
\label{subsec:aurka-beads}
To prepare artificial centrosomes, beads were coated with anti Aurora kinase A (AurkA) antibody, provided by the Mitchison lab and prepared as described before \cite{nguyen2014spatial}. The AurkA antibody recruits native Aurora kinase A from the extract to the surface of the beads, inducing microtubule nucleation. The antibody was labeled by the Antibody facility at MPI-CBG with DyLight 488. For the coating, Protein A coated Dynabeads (1.5 $\mu$g/ml suspension,  2.8 $\mu$m diameter) were washed and dispersed in a 20 $\mu$g/ml solution of the labeled AurkA antibody, and washed the next morning after overnight rotation at 4 °C.

\subsection{Extract sample preparation and imaging}
\label{subsec:sample_prep_imaging}

To induce microtubule aster formation and visualize cellular structures, the cytoplasmic extract was supplemented with artificial centrosomes (AurkA beads, prepared as described in subsection \ref{subsec:aurka-beads}) and fluorescently labeled tubulin. To prepare the reactions, 50~$\mu$l of undiluted extract was combined with 0.6~$\mu$M pig tubulin conjugated to Alexa Fluor 647 and varying dilutions of AurkA beads to achieve bead densities of roughly 100--200 beads per field of view. The mixture was assembled in an Eppendorf tube kept on ice, gently flicked several times to ensure uniform distribution of reagents, and incubated on ice for 3--5 minutes. A 4~$\mu$l aliquot of the prepared reaction was then transferred to a well of a $\mu$-Slide 8 Well high Bioinert (ibidi, 80800), a specialized imaging platform with a coating that allows high-resolution imaging of microtubules. The sample was covered with 300~$\mu$l of anti-evaporation oil (ibidi, 50051) to prevent evaporation and maintain oxygen exchange, ensuring sample viability during extended imaging sessions. 

Imaging was performed using a spinning disk confocal microscope (Olympus IX83 equipped with a CSU-W1 Yokogawa disk) connected to two Hamamatsu ORCA-Fusion BT Digital CMOS cameras (SD1), with 10x magnification. Z-stacks were acquired using a stage-top Z piezo with a spacing of 2--8~$\mu$m between planes. The time between frames ranged from 20 to 90 seconds.

\section{Experimental analysis methods}

This section describes the analysis methods used to extract quantitative measurements from the raw microscopy data. We first describe how artificial centrosomes are tracked over time to obtain bead trajectories (Sec.~\ref{subsec:tracking}). We then present the method used to reconstruct the local phase of the cell-cycle oscillator from the microtubule intensity signal (Sec.~\ref{subsec:phase_analysis}). 
We next present the analysis establishing that bead velocity is modulated by the wave phase, and how this modulation is quantified and tested for significance across datasets (Sec.~\ref{subsec:vabs_data}). 
Next, we detail how we quantify the coupling between bead-flow direction and the phase-gradient direction of the traveling wave and how this phase-flow coupling is aggregated into statistics across repeated experiments and experimental days (Sec.~\ref{subsec:flow-wave}). Finally, we describe how the growth number is estimated experimentally from independent measurements of aster polymerization speed, cell cycle period, and interparticle spacing (Sec.~\ref{subsec:growth_number_estimation}).

\subsection{Tracking of artificial centrosomes}
\label{subsec:tracking}

Particle tracking was performed using TrackMate v7 \cite{ershov2022trackmate}, a versatile Fiji/ImageJ plugin for single-particle tracking. The analysis was conducted within FIJI/ImageJ 2.7.0 \cite{schindelin2012fiji}. The raw data consisted of spinning disk confocal microscopy images (maximum intensity projections over z were used) acquired at 10x magnification, with a resolution of $1152 \times 1152$ pixels and a pixel size of $1.3 \, \mu\text{m}$. Time-lapse sequences ranged from 20 to 90 seconds.

To enhance particle visibility, a top-hat filter was applied for background subtraction. For spot detection, the Laplacian of Gaussian (LoG) detector was used, with an estimated particle diameter of $6 \, \mu\text{m}$. The detection threshold was adjusted for each sample to ensure accurate particle detection while excluding noise. Subpixel localization was enabled to improve the precision of particle position estimates. Tracking was performed using the Simple LAP Tracker algorithm, with the maximum linking distance set between $20.0 \, \mu\text{m}$ and $120.0 \, \mu\text{m}$, depending on the speed of cytoplasmic flows observed in different samples. The maximum gap-closing distance ranged from $20.0 \, \mu\text{m}$ to $40.0 \, \mu\text{m}$, and the maximum gap-closing frames were set to 2, allowing for the linking of particles across brief interruptions in detection. Tracks were filtered to retain only those with a sufficient number of detected spots.

The resulting tracks were visually inspected, and any mislinked or incomplete tracks were manually corrected using the TrackMate graphical user interface (GUI). The tracking data, including particle positions and trajectories, were exported in XML format. Custom Python (Python 3.12) scripts were used for quantitative analysis and visualization, enabling the calculation of metrics such as particle displacement, velocity, and flow dynamics.

\subsection{Phase reconstruction from microtubule intensity}
\label{subsec:phase_analysis}

\begin{figure*}[t]
\centering
\includegraphics[width=\textwidth]{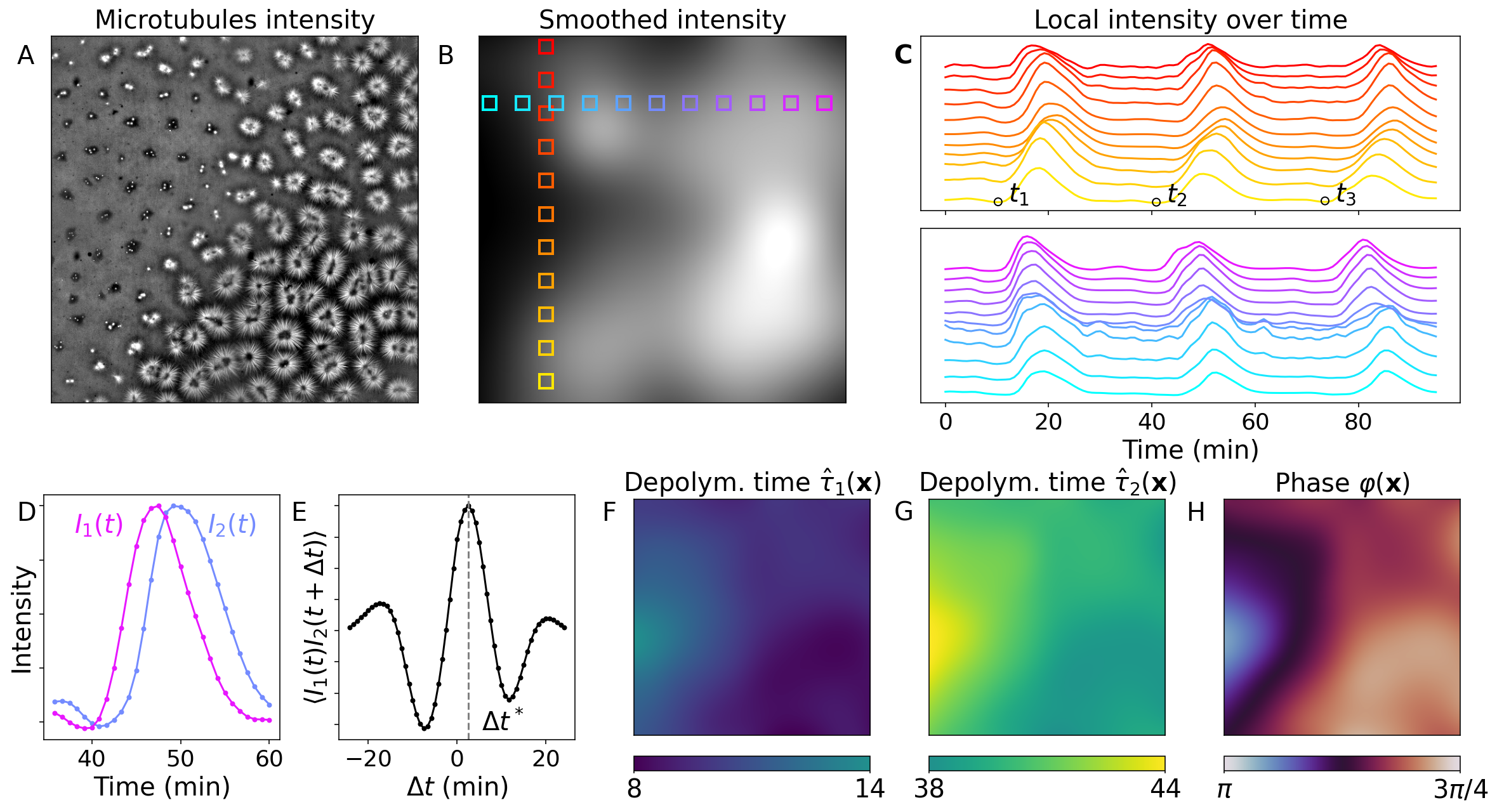}
\caption{
\textbf{Phase reconstruction from microtubule intensity dynamics.}
(A)~Raw tubulin intensity image. Individual microtubule (MT) asters are visible in polymerized regions. 
(B)~The same image after spatial smoothing with a Gaussian filter, which suppresses intensity variations associated with individual asters while preserving the large-scale wave structure.
(C)~Temporal intensity traces extracted from selected pixels along a horizontal line (top) and a vertical column (bottom). The colors of the traces correspond to the colored markers shown in panel (B), illustrating the progressive temporal shift of the oscillations across space.
(D)~Illustration of the relative delay between two neighboring pixels. The MT intensity signals \(I_1(t)\) (pink) and \(I_2(t)\) (blue) are shifted in time owing to wave propagation. 
(E)~Determination of the optimal delay \(\Delta t^*\) from the temporal cross-correlation between \(I_1(t)\) and \(I_2(t)\). The delay is defined as the lag at which the cross-correlation reaches its maximum.
(F)~Reconstructed depolymerization-time map \(\hat{\tau}_1(\mathbf{x})=\tau_1(\mathbf{x})+t_1\) for one oscillation cycle. The relative delay field \(\tau_1(\mathbf{x})\) is obtained by combining all pairwise delay measurements through a global least-squares optimization, and \(t_1\) denotes the depolymerization time of the reference pixel. 
(G)~Depolymerization-time map \(\hat{\tau}_2(\mathbf{x})=\tau_2(\mathbf{x})+t_2\) for the subsequent oscillation cycle, reconstructed using the same procedure.
(F--G)~The colorbar indicates the depolymerization time in minutes. 
(H)~Construction of the phase field from the two successive depolymerization-time maps. The local phase is defined by the normalized time elapsed between \(\hat{\tau}_1(\mathbf{x})\) and \(\hat{\tau}_2(\mathbf{x})\), yielding a continuous phase variable \(\varphi(\mathbf{x},t)\in[0,2\pi)\).
}
\label{fig:phase_analysis}
\end{figure*}

To quantify the local state of the cell-cycle oscillator, we reconstruct a phase field
\(\varphi(\mathbf{x},t)\) from the microtubule (MT) intensity signal.
Because direct measurements of Cdk1 activity are not available in these experiments, the MT intensity serves as an indirect reporter of the oscillator dynamics. The method relies on determining the timing of microtubule depolymerization events across the extract and using these events as phase markers.

Figure~\ref{fig:phase_analysis} illustrates the complete reconstruction procedure.

First, the raw MT intensity field \(I(\mathbf{x},t)\) (Fig.~\ref{fig:phase_analysis}A) is smoothed in space using a Gaussian filter (Fig.~\ref{fig:phase_analysis}B). This filtering suppresses intensity fluctuations associated with individual asters and reveals the large-scale structure of the propagating mitotic waves. For each pixel \(\mathbf{x}\), the resulting intensity signal \(I(\mathbf{x},t)\) exhibits periodic oscillations associated with successive cycles of microtubule polymerization and depolymerization (Fig.~\ref{fig:phase_analysis}C). In some cases, prior to smoothing, we square the intensity signal to emphasize large oscillations and reduce the relative contribution of small ones.

To determine the relative timing of the oscillations, we compare the intensity traces of neighboring pixels. Consider two neighboring pixels with signals \(I_1(t)\) and \(I_2(t)\). Because the mitotic wave propagates through the extract, the two signals are typically shifted in time with respect to one another (Fig.~\ref{fig:phase_analysis}D). The relative delay \(\Delta t^*\) is obtained by maximizing their temporal cross-correlation,
\begin{equation}
C_{12}(\Delta t)
=
\int I_1(t)\,I_2(t+\Delta t)\,dt ,
\end{equation}
such that
\begin{equation}
\Delta t^*
=
\underset{\Delta t}{\mathrm{arg\,max}}
\; C_{12}(\Delta t).
\end{equation}
At the optimal delay \(\Delta t^*\), the two oscillatory signals are maximally aligned (Fig.~\ref{fig:phase_analysis}E).

This procedure is repeated for all pairs of neighboring pixels, yielding a set of measured pairwise delays \(\Delta t^*_{ij}\). These local measurements are then combined into a global delay map \(\tau(\mathbf{x})\) by solving a least-squares optimization problem. Denoting by \(i\) and \(j\) neighboring pixels, we determine \(\tau\) by minimizing
\begin{equation}
\mathcal{L}
=
\sum_{\langle i,j\rangle}
\left(
\tau_j-\tau_i-\Delta t^*_{ij}
\right)^2 ,
\end{equation}
subject to the constraint that the delay of a reference pixel is fixed to zero. The resulting field \(\tau(\mathbf{x})\) provides the relative timing of the oscillation throughout the extract.

To convert these relative timings into absolute depolymerization times, we identify a reference pixel and determine the times \(t_1\) and \(t_2\) at which its intensity reaches successive minima. These minima correspond to microtubule depolymerization events and define the beginnings of two consecutive oscillation cycles. Adding the reference times to the reconstructed delay maps yields the spatially resolved depolymerization times
\begin{equation}
\hat{\tau}_1(\mathbf{x})
=
\tau_1(\mathbf{x}) + t_1,
\end{equation}
and
\begin{equation}
\hat{\tau}_2(\mathbf{x})
=
\tau_2(\mathbf{x}) + t_2,
\end{equation}
for two consecutive cycles (Fig.~\ref{fig:phase_analysis}F--G).

Finally, the oscillator phase is defined as the fraction of elapsed time between successive depolymerization events. For a time \(t\) satisfying
\(
\hat{\tau}_1(\mathbf{x})
\leq
t
<
\hat{\tau}_2(\mathbf{x})
\),
the phase is given by
\begin{equation}
\varphi(\mathbf{x},t)
=
2\pi
\,
\frac{
t-\hat{\tau}_1(\mathbf{x})
}{
\hat{\tau}_2(\mathbf{x})
-\hat{\tau}_1(\mathbf{x})
},
\end{equation}
with \(\varphi\in[0,2\pi)\) (Fig.~\ref{fig:phase_analysis}H). By construction, \(\varphi=0\) immediately after depolymerization and increases continuously until the next depolymerization event, where it resets to zero. We systematically verify by visual inspection that the resulting phase pattern is consistent with the wave pattern observed in the raw intensity data.

This approach provides a robust estimate of the local oscillator phase directly from the MT intensity dynamics. Unlike standard phase-extraction techniques that require direct measurements of biochemical activity, it only assumes the existence of reproducible depolymerization events and therefore remains applicable in systems where molecular reporters of oscillator activity are unavailable \cite{tan2020topological}.

\subsection{Statistics over several experiments for the phase modulated speed}
\label{subsec:vabs_data}

\begin{figure}
    \centering
    \includegraphics[width=\linewidth]{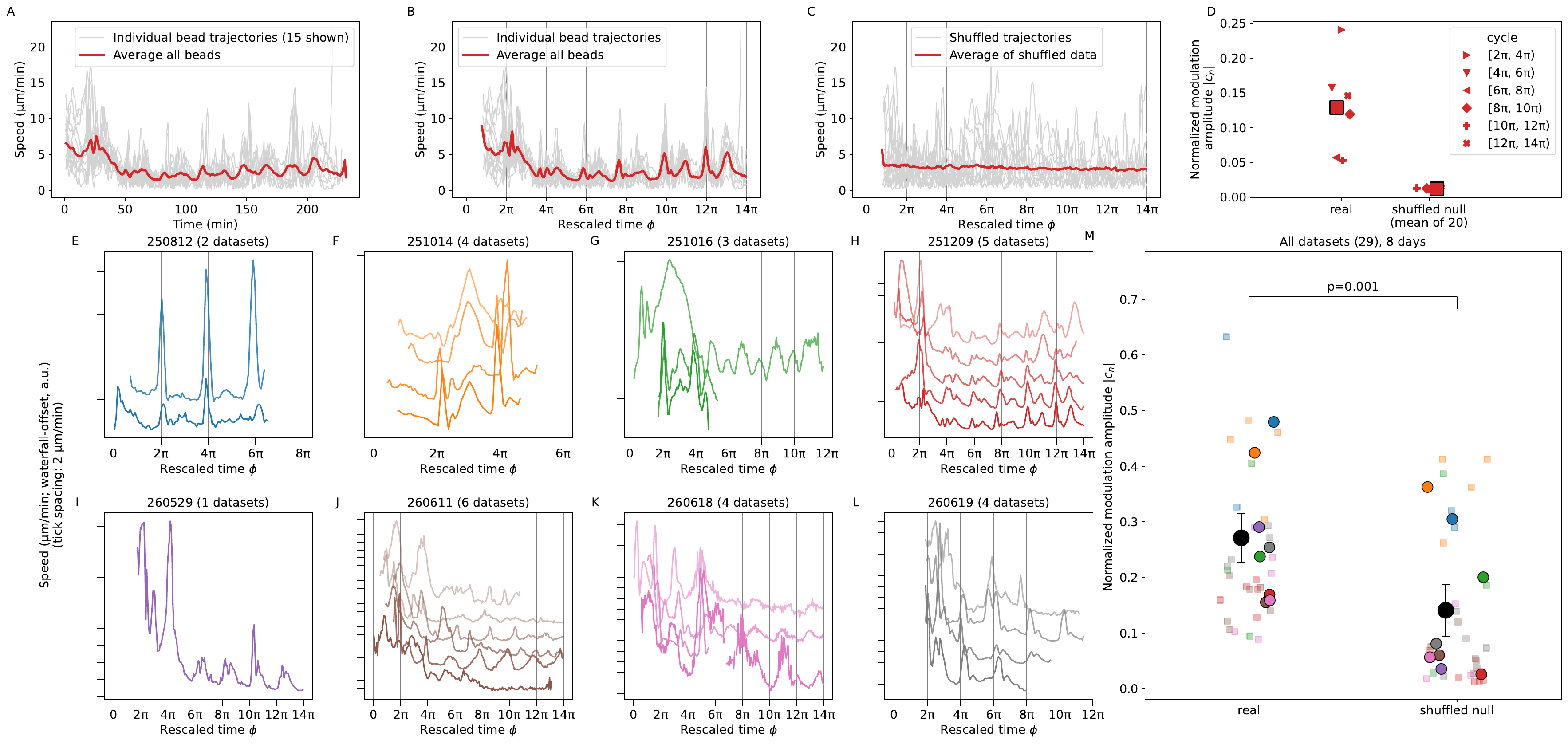}
    \caption{
\textbf{The speed of the beads is modulated by the phase}\\
(A)~Average bead speed ($\mu$m/min) in real time (min); thin grey lines: individual bead trajectories; thick colored line: average over all beads.
(B)~Same dataset, average bead speed vs.\ each bead's own rescaled time $\phi$ (Eq.~\ref{eq:unwrapped time}); grey/colored lines as in (A).
(C)~Same as (B), but each trajectory's speed values are independently shuffled by a random circular roll before averaging (one realization of the null model).
(D)~Per-cycle modulation amplitude $|c_n|$ for this dataset, real (left) vs.\ shuffled null (right, mean of $20$ independent rolls). Small markers: one point per wave cycle, marker shape coding that cycle's phase interval (legend, units of $\pi$); large square: average across cycles.
(E--L)~One panel per day, each showing every dataset from that day's own mean speed-vs-rescaled-time curve. When a day has more than one dataset, curves after the first are vertically offset, to separate overlapping curves. Tick marks are spaced at a fixed 2~$\mu$m/min interval (shared $y$-axis label at left) so tick-to-tick height is comparable across panels.
(M)~Modulation amplitude $|c_n|$, real vs.\ shuffled null, summarizing every dataset/day. Faded squares: per-dataset values (averaged over that dataset's own cycles). Larger circles: per-day averages, colored by day. Black circle with error bar: average across days with standard error. Bracket and $p$-value: paired $t$-test between real and shuffled-null values across days.
}
\label{fig:vabs_phase}
\end{figure}

Because the wave propagates in space, it reaches beads at different positions with different delays; a bead's local rescaled time — the time elapsed since the wave phase last passed that position — accounts for this delay. We therefore ask whether the speed peaks observed at different times and positions collapse onto a single, shared profile once each trajectory's speed is expressed as a function of its own rescaled time, rather than of absolute time. To confirm that any resulting peak reflects genuine phase-locking rather than an artifact of binning or averaging, we validate it here against a shuffled null model at three levels: individual trajectories, individual datasets, and across experimental days.

\paragraph*{Phase-resolved speed profile.}
For each bead trajectory, we computed the local rescaled time
\begin{equation}
\phi(t) = 2\pi n + \varphi(t), 
\label{eq:unwrapped time}
\end{equation}
where $\varphi(t) \in [0, 2\pi)$ is the wrapped phase extracted from the intensity signal at the bead's position and $n$ counts the number of elapsed wave periods following the method described in Sec.~\ref{subsec:phase_analysis}. Plotting mean bead speed as a function of real time (Fig.~\ref{fig:vabs_phase}A) shows some shared temporal structure across beads, which is enhanced when re-expressing the same data as a function of each bead's own rescaled time (Fig.~\ref{fig:vabs_phase}B). It reveals two clear, reproducible peak shared across trajectories, indicating that speed is controlled by wave phase.

\paragraph*{Null model.}
To test whether this peak reflects genuine phase-locking rather than a binning artifact, we constructed a null model in which each trajectory's speed time series was circularly shifted by an independent random offset before recomputing the phase-averaged profile (Fig.~\ref{fig:vabs_phase}C). A circular shift was used to preserve each trajectory's own autocorrelation structure (e.g., frame-to-frame persistence in speed) while fully destroying its temporal alignment with phase; this isolates the phase-coupling effect from any secondary structure intrinsic to the speed trace itself. Under this null, the peaks observed in Fig.~\ref{fig:vabs_phase}B disappear (Fig.~\ref{fig:vabs_phase}C), consistent with the peak reflecting a genuine phase relationship rather than an artifact of the averaging or binning procedure.

\paragraph*{Quantifying modulation amplitude, $|c_n|$.}
To quantify this effect per wave cycle and enable statistical comparison across cycles, datasets, and days, we defined, for each trajectory and each wave cycle $n$, a complex modulation coefficient
\begin{equation}
c_n = \sum_b \tilde v_n(b)\, e^{-i\phi_b},
\end{equation}
where $\tilde v_n(b)$ is the trajectory's speed averaged into phase bin $b$ of cycle $n$, normalized by that trajectory's own mean speed over cycle $n$ (removing baseline speed differences between beads and between cycles), and $\phi_b$ is the bin's phase center. The magnitude $|c_n|$ quantifies the depth of phase modulation within a single cycle, independent of the absolute phase at which the peak occurs, which allows comparison across cycles and trajectories even when the precise phase of peak speed is not perfectly stationary. Fig.~\ref{fig:vabs_phase}D shows $|c_n|$ for one representative dataset, with each small marker corresponding to one wave cycle (marker shape indicating the cycle's approximate phase interval) and the large square marking the average across cycles; the corresponding null value (right) was obtained by averaging $|c_n|$ over $N_\mathrm{null\,reps}$ independent circular-shift realizations per cycle. The real modulation amplitude clearly exceeds the null in this example.

\paragraph*{Consistency across datasets and days.}
Fig.~\ref{fig:vabs_phase}E--L show the phase-resolved speed profile (as in Fig.~\ref{fig:vabs_phase}B) for every dataset, grouped by the day on which it was acquired; within a day, curves from additional datasets are vertically offset for visual clarity, and the shared $y$-axis tick spacing (2~$\mu$m/min) allows the modulation amplitude to be compared directly across panels. Peaks of comparable shape and amplitude are observed across datasets and across days, indicating that the phase-speed coupling is a reproducible feature of the system rather than being specific to a single experiment.

\paragraph*{Statistical significance across days.}
To test significance while respecting the non-independence of datasets acquired on the same day, we aggregated $|c_n|$ hierarchically: cycles were averaged within each dataset (faded squares, Fig.~\ref{fig:vabs_phase}M), datasets were averaged within each day (larger circles, colored by day), and the resulting per-day values were compared between the real and shuffled-null conditions using a paired $t$-test, with each day contributing one paired observation (real vs.\ null). This hierarchical averaging ensures that the reported $p$-value reflects the number of independent experimental days rather than the much larger (and non-independent) number of underlying cycles or trajectories. The real modulation amplitude was significantly higher than the null across days (Fig.~\ref{fig:vabs_phase}M, bracket and $p$-value), and the black circle with error bar (mean and standard error across days) summarizes the overall effect size. Only wave cycles in which more than 10 trajectories contributed a valid speed estimate were included in this analysis, to avoid noisy $|c_n|$ estimates from sparsely sampled cycles.

\subsection{Statistics for the phase-flow coupling}
\label{subsec:flow-wave}

Here, we explain how we establish the panel of Fig.~1E of the main text. 

\begin{figure}
    \centering
    \includegraphics[width=\linewidth]{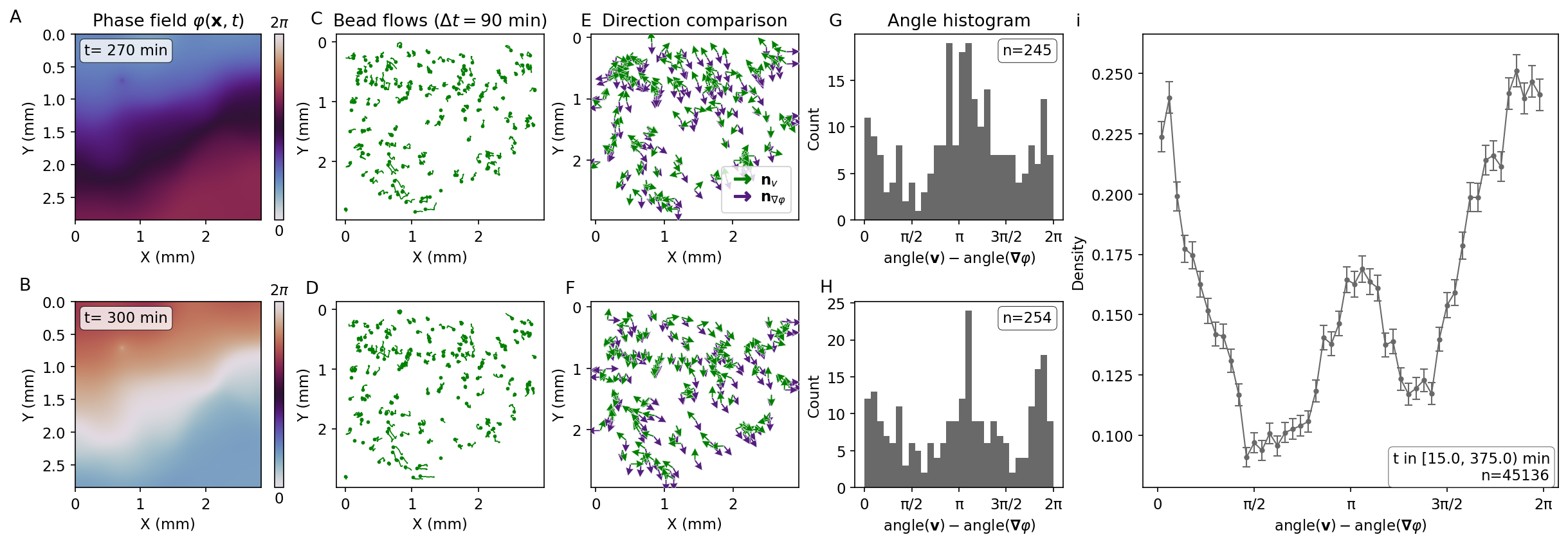}
    \caption{\textbf{Local bead-flow direction tracks the phase gradient of the traveling wave.}
    (A--B)~Reconstructed phase field $\varphi(\mathbf{x},t)$ from the microtubule intensity following the method described in Sec.\ref{subsec:phase_analysis} at the indicated time.
    (C--D)~Corresponding bead trajectories (trailing window ending at the time indicated in (A--B)) measured independently.
    (E--F)~Local bead-velocity direction ($\mathbf{n}_v$, green) and local phase-gradient direction
    ($\mathbf{n}_{\nabla\varphi}$, purple) at each bead (direction
    only; arrow length is not meaningful).
    (C--F)~Only $60\%$ of the bead trajectories and bead directions are shown for clarity.
    (G--H)~Distribution of the angle between $\mathbf{n}_v$ and $\mathbf{n}_{\nabla\varphi}$ at that
    timepoint for all beads.
    (I)~Same distribution pooled over the whole analyzed time window, plotted as a density with
    counting-statistics error bars. Concentration near $0$ and $\pi/2$ indicates that bead motion locally follows the
    direction of wave propagation.}
    \label{fig:phase_flow_angle}
\end{figure}

\paragraph*{Individual measure of the wave-flow alignment}
For each frame of each dataset, we separately compute the phase field from the microtubule intensity (Sec.~\ref{subsec:phase_analysis}) (Fig.~\ref{fig:phase_flow_angle}A--B) and the velocity of the beads from their trajectories (Fig.~\ref{fig:phase_flow_angle}C--D). 

The local phase-gradient direction at each pixel is obtained from a structure-tensor analysis of the phase field, which yields a locally averaged, noise-robust gradient direction even where the raw pixel-to-pixel gradient is noisy. The gradient direction is displayed by the purple arrows in Fig.~\ref{fig:phase_flow_angle}E--F. 

The bead velocity is computed by a central finite difference with \(dt = 2\) frames; when a trajectory has no detection at exactly the time needed, the missing point is reconstructed by linear interpolation between its surrounding, existing detections. The flow direction is shown by the green arrows in Fig.~\ref{fig:phase_flow_angle}E--F. 

At each position where a bead is recorded, we then compute the angle difference between the phase-gradient direction and the bead velocity direction at that position (Fig.~\ref{fig:phase_flow_angle}E--F). We compute the distribution of this angle difference at each frame (Fig.~\ref{fig:phase_flow_angle}G--H), and average it over the whole duration of the experiment (Fig.~\ref{fig:phase_flow_angle}I).

\begin{figure}
    \centering
    \includegraphics[width=0.8\linewidth]{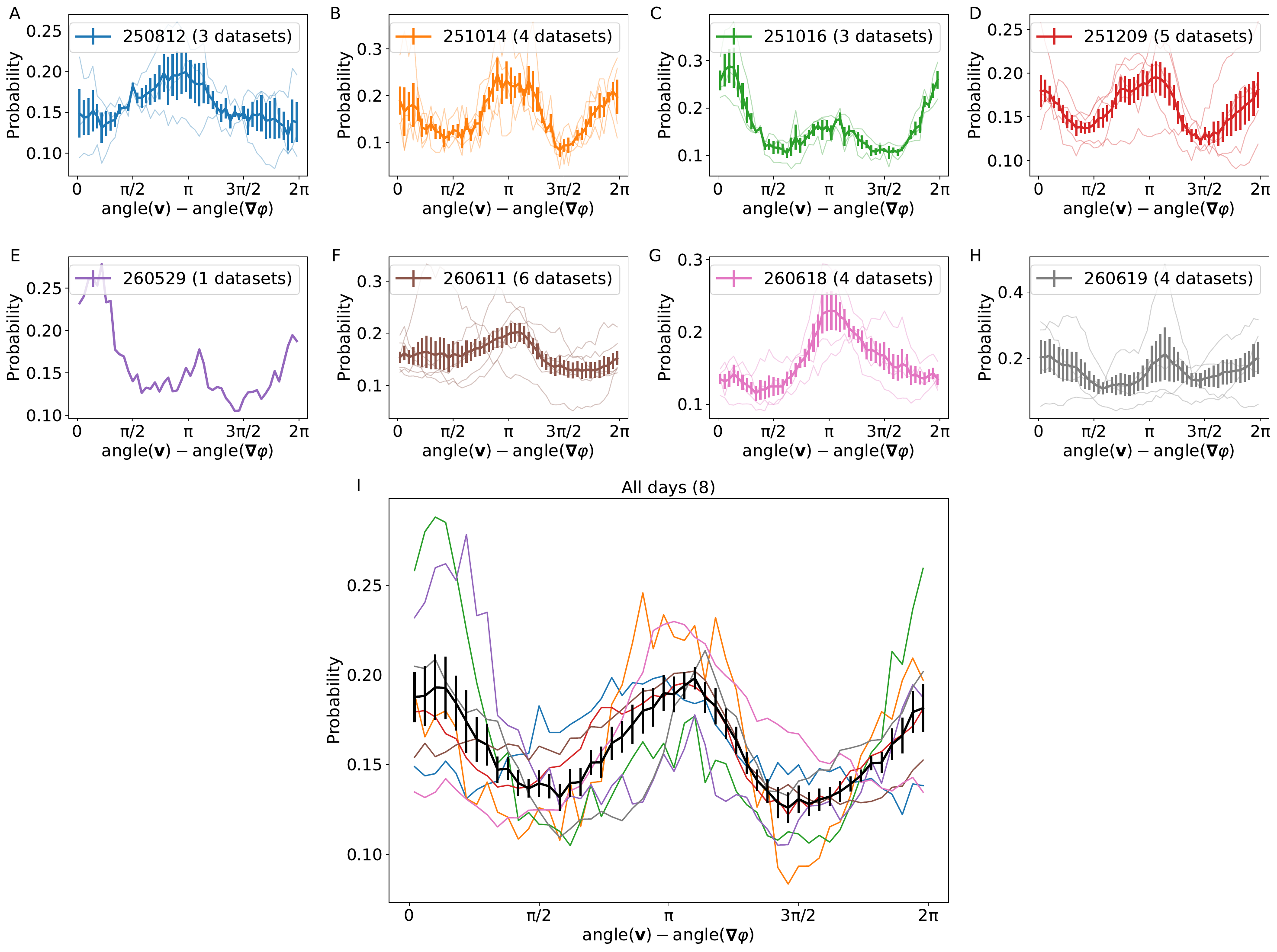}
    \caption{\textbf{Bead flows align with the wave direction.} \\
    (A--H)~Distribution of the angular difference between bead velocity ($\mathbf{v}$) and the wave-propagation direction ($\mathbf{\nabla}\varphi$), for each of 8 independent days (extract preparations). Thin lines show individual replicates; the thick line is the daily average $\pm$ standard error. 
    (I)~Per-day averages from (A--H), overlaid (colors as in A--H), with the overall average across all 8 days in black. Error bars show the standard error. This average is used in Fig.~1E of the main text. 
    }
    \label{fig:data-angle}
\end{figure}

\paragraph*{Statistical significance across days.}
We now describe how we compute statistics across repeated experiments. On each of several days, we perform multiple independent imaging acquisitions of the same extract preparation, each in a different droplet. For each dataset, we perform the phase analysis and compute the flow--wave coupling statistics as described in Secs.~\ref{subsec:phase_analysis} and above. Individual results are shown in Fig.~\ref{fig:data-angle}(A--H). We then averaged the angular-difference density distributions within each day, and subsequently across all days (Fig.~\ref{fig:data-angle}I).

\subsection{Estimation of the growth number from experimental measurements}
\label{subsec:growth_number_estimation}
 
To estimate the growth number \Gr~(Eq.~2 of the main text) experimentally, we independently measured its three constituent quantities from 4 imaging experiments: the aster polymerization velocity $v_p$, the cell cycle period $T_\mathrm{osc}$, and the characteristic interparticle spacing $\ell_0$. We used a representative dataset of 4 imaging experiments (averaged per experiment day to account for cytoplasmic extract variability) instead of the full dataset used throughout this study, due to the manual nature of the measurements.
 
\paragraph*{Artificial centrosomes detection.}
Unlike the trajectory-based tracking described in Sec.~\ref{subsec:tracking}, which links bead positions across frames to compute velocities, the spatial statistics below only require instantaneous bead positions and therefore do not require linking. For each timelapse, bead positions were detected independently in every frame using TrackMate's Laplacian-of-Gaussian (LoG) detector~\cite{ershov2022trackmate}, applied to maximum-intensity-projected images of the AurkA beads, with no frame-to-frame tracking step. Detection parameters (blob radius of $6.0~\mu m$, with a quality threshold of $10.0$) were tuned interactively on representative movies and then applied uniformly in batch across the dataset. Multiple consecutive periods, each corresponding to a successive cell-cycle oscillation captured for the same field of view, were pooled as replicates of the same experimental position for averaging purposes (see below).
 
\paragraph*{Microtubule aster clustering.}
Because closely spaced beads can nucleate microtubules that grow into a single shared aster rather than remaining as independent asters, bead detections separated by less than $30~\mu\text{m}$ were merged into a single aster position prior to spatial analysis. This threshold was chosen based on direct visual inspection of the timelapses: beads closer than $30~\mu\text{m}$ consistently gave rise to microtubule networks that overlapped and grew as a single visually indistinguishable aster, whereas beads farther apart remained visually separable throughout growth. This value is small relative to the typical aster radius at impingement ($126\pm22~\mu\text{m}$, mean $\pm$ SD, measured from the aster polymerization speed dataset described below), such that beads within this distance are expected to share a continuous microtubule network well before either aster approaches its full size. The threshold is also far larger than the sub-micrometer localization precision of the subpixel LoG detector, so detection noise is not expected to drive spurious merging. Merging was performed per frame using a transitive (single-linkage) rule: detections were connected in a graph if their pairwise distance fell below this threshold, and the position of each resulting connected component was taken as the centroid of its constituent detections. All spatial statistics below were computed on the merged aster positions.

\paragraph*{Particle density and characteristic interparticle spacing $\ell_0$.}
For each frame, we computed the particle density as the number of detected aster positions divided by the imaged field-of-view area,
\begin{equation}
\rho(t) = \frac{N(t)}{A},
\end{equation}
where $N(t)$ is the number of asters detected at time $t$ and $A = (1152\times1.3~\mu\text{m})\times(1152\times1.3~\mu\text{m}) \approx 2.24~\text{mm}^2$ is the imaged field-of-view area, determined by the camera resolution and pixel size. Density was averaged over 2--3 cell cycle periods for the same imaging position (equal weight per period), then across positions imaged on the same day (equal weight per position), and finally across days to obtain the grand-average estimate; standard errors were propagated at each averaging step. The characteristic interparticle spacing entering the growth number, $\ell_0 = L/\sqrt{N}$ (main text, Eq.~2), is, for a field of view of area $A=L^2$ containing $N$ particles, equivalent to $\ell_0 = 1/\sqrt{\rho}$. We therefore estimated $\ell_0$ directly from the day-averaged density. Because $\ell_0 \propto \rho^{-1/2}$, its relative standard error is half that of the underlying density measurement,
\begin{equation}
\frac{\delta \ell_0}{\ell_0} = \frac{1}{2}\,\frac{\delta \rho}{\rho} .
\label{eq:si_ell0_error_propagation}
\end{equation}
Across the full dataset of 4 days, the grand-average characteristic spacing is $\ell_0 = 155.2 \pm 19.4~\mu\text{m}$ (mean $\pm$ SEM across days; SD $= 38.8~\mu\text{m}$), with day-level averages ranging from 105.8 to 194.3~$\mu$m. (Table~\ref{tab:si_growth_number_days}).

\paragraph*{Microtubule aster polymerization speed.}
The radial growth speed $v_p$ of individual asters was measured manually from the same timelapse movies. For 2--3 representative asters per movie, we recorded the aster radius (distance from the bead center to the visible microtubule tip) at its minimal size immediately following depolymerization, and again at the frame at which the growing aster first made contact with a neighboring aster (impingement), together with the number of elapsed frames between the two timepoints. The average polymerization speed for each aster was computed as $v_p = (r_\mathrm{end}-r_\mathrm{start})/\Delta t$, where $\Delta t$ is the elapsed time (frame interval multiplied by elapsed frames). Individual aster speeds were averaged within each file, then across period files within each position, then across positions imaged on the same day, and finally across days, following the same hierarchical averaging scheme used for $\ell_0$. Measuring growth up to the point of impingement, rather than over a fixed time window, provides a physically motivated and reproducible stopping criterion; because impingement occurs earlier in more densely packed regions, the measured speed reflects growth under the local mechanical confinement set by neighboring asters, consistent with the environment-dependent interpretation of the growth number. Across days, the grand-average polymerization speed is $v_p = 15.43 \pm 0.97~\mu\text{m/min}$ (mean $\pm$ SEM; SD $=1.93~\mu\text{m/min}$).
 
\paragraph*{Cell cycle period.}
The cell cycle period $T_\mathrm{osc}$ was obtained independently for each movie from the phase reconstruction analysis described in Sec.~\ref{subsec:phase_analysis}, as the time between successive depolymerization events identified from the microtubule intensity field. Across days, the grand-average cell-cycle period is $T_\mathrm{osc} = 39.8 \pm 6.0$~min (mean $\pm$ SEM; SD $=12.0$~min).
 
\paragraph*{Error propagation and growth number estimation.}
For each imaging position, we combined the position-averaged estimates of $v_p$, $T_\mathrm{osc}$, and $\ell_0$ to compute the experimental growth number. For clarity purposes in this section we use the following notation for the growth number: $\mathrm{Gr}_\mathrm{exp} = v_p T_\mathrm{osc}/\ell_0$. Because these three quantities were measured independently (from manual aster measurements, particle detection, and phase analysis, respectively), we propagated their standard errors in quadrature assuming independence,
\begin{equation}
\left(\frac{\delta \mathrm{Gr}_\mathrm{exp}}{\mathrm{Gr}_\mathrm{exp}}\right)^2
=
\left(\frac{\delta v_p}{v_p}\right)^2
+
\left(\frac{\delta T_\mathrm{osc}}{T_\mathrm{osc}}\right)^2
+
\left(\frac{\delta \ell_0}{\ell_0}\right)^2 ,
\label{eq:si_gr_error_propagation}
\end{equation}
where each $\delta$ denotes the standard error of the mean of the corresponding position-averaged quantity. Position-level $\mathrm{Gr}_\mathrm{exp}$ values were averaged across positions imaged on the same day (equal weight per position), and these day-level averages were then averaged across days to obtain the final grand-average estimate. Over 4 experiment days (9 imaging positions in total), the experimental growth number is $\mathrm{Gr}_\mathrm{exp} = 4.3~\pm 1.2~(mean~\pm~SEM~across~days;~SD = 2.37)$. For reference, the flat average across all positions without day-level grouping is  $4.13 \pm 0.7~(n=9~\mathrm{positions})$. Table~\ref{tab:si_growth_number_days} reports $v_p$, $T_\mathrm{osc}$, $\ell_0$, and the resulting $\mathrm{Gr}_\mathrm{exp}$ averaged per day.

\begin{table}
\centering
\caption{Growth-number components averaged across positions imaged on the same day (equal weight per position). $N_\mathrm{pos}$ shows how many positions were averaged per day. Values are mean $\pm$ SEM.}
\label{tab:si_growth_number_days}
\begin{tabular}{lccccc}
\hline\hline
Experiment day & $N_\mathrm{pos}$ & $v_p$ ($\mu$m/min) & $T_\mathrm{osc}$ (min) & $\ell_0$ ($\mu$m) & $\mathrm{Gr}_\mathrm{exp}$ \\
\hline
20251209 & 3 & 14.82 $\pm$ 0.51 & 36.17 $\pm$ 0.73 & 194.3 $\pm$ 13.69 & 2.77 $\pm$ 0.07 \\
20260611 & 2 & 13.13 $\pm$ 1.01 & 48.15 $\pm$ 10.05 & 144.57 $\pm$ 2.46 & 4.31 $\pm$ 0.65 \\
20260618 & 2 & 17.69 $\pm$ 0.32 & 24.45 $\pm$ 0.95 & 176.28 $\pm$ 1.54 & 2.45 $\pm$ 0.07 \\
20260619 & 2 & 16.09 $\pm$ 0.38 & 50.30 $\pm$ 2.50 & 105.8 $\pm$ 0.67 & 7.63 $\pm$ 0.15 \\
\hline\hline
\end{tabular}
\end{table}

\section{Simulation methods and analysis}

In this section, we detail the simulation methods and numerical procedures used to characterize spatial and temporal organization in the system. In Sec.~\ref{sec:simulation_methods}, we describe the simulation setup, including initial conditions, boundary conditions, and the numerical integration scheme. In Sec.~\ref{sec:si_correlation_length}, we describe how the discrete particle phases are interpolated onto a continuous field $\varphi(\mathbf{r})$ and how its spatial correlation length $\xi$ is extracted; the same interpolation procedure is used to construct the coarse-grained velocity fields shown in the main text. In Sec.~\ref{sec:si_connectivity_correlation}, we describe how the particle adjacency matrix $M_{ij}$ is constructed from a Delaunay triangulation and how its temporal autocorrelation is used to extract the characteristic neighbor exchange time $\tau$.

\subsection{Simulation methods}
\label{sec:simulation_methods}

We performed numerical simulations of the coupled mechanochemical dynamics defined by Eqs.~(1), (3), and (4) in a two-dimensional square domain with periodic boundary conditions. The simulation were coded in Julia. Unless stated otherwise, particles were initialized on a regular lattice with small random positional perturbations to avoid artificial overlap symmetries, while particle phases were uniformly distributed in the interval $[0,2\pi)$. The system was then evolved using overdamped dynamics with discrete particle depolymerization events occurring (size reset to $\sigma_0)$ when the phase reached $2\pi$.

To efficiently compute short-range mechanical interactions between overlapping particles, we implemented a cell-list algorithm combined with Verlet neighbor lists. Neighbor lists were updated periodically throughout the simulation to account for particle motion and size changes. Numerical integration was performed with a fixed time step sufficiently small to resolve both the phase dynamics and the rapid depolymerization events.

For the simulations shown in Fig.~2, particles were not initialized with random phases. Instead, we imposed a prescribed spatial phase gradient corresponding either to a planar traveling wave or to a phase field containing a topological defect, depending on the geometry considered. For sufficiently large synchronization strength $\epsilon$, these imposed phase profiles remained phase-locked throughout the simulation and generated stable propagating waves.
For the phase diagram of Fig.~3, the parameters are $\omega=1$, $\epsilon=0.5$, $D_\varphi/\epsilon=0.01$, $\mu U_0=2$, $r_0/\ell_0=0.05$.

\subsection{Phase field interpolation and correlation length}
\label{sec:si_correlation_length}

To characterize the spatial organization of the phase field $\varphi$, we compute its two-point spatial correlation function and extract a correlation length $\xi$ through the following procedure, applied to the final simulation state.

\paragraph*{Interpolation of the phase field onto a regular grid}

The raw simulation output consists of an off-lattice point cloud of $N$ particles with positions $(x_i, y_i)$ and an associated scalar phase $\varphi_i \in [0,2\pi)$, defined in a periodic (or reflective) domain of size $L \times L$. The phase is represented as a complex unit vector $\psi_i = e^{i \varphi_i}$, which avoids the discontinuity of averaging a circular quantity directly.

The domain is discretized into a regular grid of $n \times n$ bins, with a grid resolution twice as fine as the mean interparticle spacing $\ell_0$. Particle positions are binned onto this grid, and the complex phase is averaged within each occupied bin:
\begin{equation}
    \psi(\mathbf{r}_{jk}) = \frac{1}{n_{jk}} \sum_{i \in \text{bin } jk} \psi_i,
\end{equation}
where $n_{jk}$ is the number of particles in bin $(j,k)$; empty bins are set to $\psi = 0$. The real and imaginary parts of $\psi(\mathbf{r})$ are then smoothed independently with a Gaussian filter of standard deviation $\sigma_{g} = 1$ grid unit, into $\psi_{\sigma}(\mathbf{r})$ using periodic boundary conditions. The coarse-grained phase field is finally recovered as
$\varphi(\mathbf{r}) = \arg\left[\psi_\sigma(\mathbf{r})\right]$. 
Phase fields for the steady states of the different systems are shown in Fig.~1H and Fig.~3A of the main text. The velocity fields shown in Fig.~1H, Fig.~2C, and Fig.~2E are obtained with the same binning and Gaussian smoothing procedure, applied directly to the Cartesian components $(v_x, v_y)$ of the finite-difference velocity of each bead (computed from bead positions at consecutive saved time steps), rather than to a complex phase; no argument/phase-wrapping step is involved in this case.

\paragraph*{Radial correlation function}

From the interpolated phase field $\varphi(\mathbf{r})$ we define the associated planar unit vector field $\boldsymbol{\phi}(\mathbf{r}) = \big(\cos\varphi(\mathbf{r}),\, \sin\varphi(\mathbf{r})\big)$, and compute its two-point autocorrelation using the Wiener--Khinchin theorem. The correlation map is $C(\mathbf{r}) = \langle \boldsymbol{\phi}(\mathbf{r}_0)\cdot\boldsymbol{\phi}(\mathbf{r}_0+\mathbf{r})\rangle_{\mathbf{r}_0}$, computed efficiently via FFT under the assumption of periodicity of the grid.

The correlation map is then radially averaged. Distances from the center are binned into $n_b = n/2$ linearly spaced bins spanning $r \in [0, r_{\max}]$, with $r_{\max} = L/2$, and the correlation function in each bin is the mean of $C(\mathbf{r})$ over all pixels falling within that radial shell.

\paragraph*{Extraction of the correlation length}

The radial correlation function $C(r)$ is fitted by nonlinear least squares to an exponentially decaying form with a constant plateau,
\begin{equation}
    C(r) = C_0 \, e^{-r/\xi} + C_\infty,
    \label{eq:exp_fit}
\end{equation}
where $\xi$ is the correlation length, $C_0$ sets the amplitude of the decaying part, and $C_\infty$ accounts for a residual long-range offset (e.g., due to finite-size or global order effects). The phase diagram of $\xi$ for varying $\epsilon$ and \Gr is shown in Fig. 3B of the main text.

\subsection{Adjacency matrix and neighbor exchange correlation time}
\label{sec:si_connectivity_correlation}

To quantify particle mixing, we construct at each saved time step the adjacency matrix $M_{ij}$ of the instantaneous particle configuration and analyze its temporal autocorrelation $C(\Delta t)$, from which we extract a neighbor exchange (mixing) time $\tau$.

\paragraph*{Construction of the adjacency matrix}
At each time $t$, particle positions $(x_i, y_i)$ are triangulated using a Delaunay triangulation. To account for periodic boundary conditions, particles located near the domain edges are duplicated on the opposite side of the box prior to triangulation, and any resulting edge involving such a periodic image is mapped back to the index of the corresponding real particle. Two particles $i,j$ are considered connected, $M_{ij}(t) = 1$, if they share an edge of the triangulation; $M_{ij}(t)=0$ otherwise. The matrix $M(t)$ is symmetric, and configurations are saved at intervals $\delta t = T_\mathrm{osc}$.

\paragraph*{Temporal autocorrelation}
To avoid contamination from the initial transient, the autocorrelation is computed only over configurations saved after $t=60\,T_\mathrm{osc}$. For a given lag $\Delta t = k\,\delta t$ ($k=0,1,2,\dots$), the similarity between adjacency matrices separated by $\Delta t$ is quantified through the overlap of their edges,
\begin{equation}
    C(\Delta t) = \frac{1}{T-k}\sum_{t=60\,T_\mathrm{osc}}^{T-k-1} \sum_{i,j} M_{ij}(t)\, M_{ij}(t+\Delta t),
\end{equation}
normalized by a global constant equal to the total number of edges accumulated over the retained trajectory, $\sum_{t \geq 60\,T_\mathrm{osc}} \sum_{i,j} M_{ij}(t)^2$, so that $C(\Delta t)$ decreases from its initial value as neighbor relations decorrelate. Time is expressed in units of the oscillation period $T_\mathrm{osc}$.

\paragraph*{Extraction of the mixing time}
As for the phase correlation length (Sec.~\ref{sec:si_correlation_length}), the autocorrelation $C(\Delta t)$ is fitted by nonlinear least squares to an exponentially decaying form with a constant plateau,
\begin{equation}
    C(\Delta t) = C_0 \, e^{-\Delta t/\tau} + C_\infty,
\end{equation}
where $\tau$ is the characteristic neighbor exchange (mixing) time, and $C_\infty$ accounts for a residual long-time offset in the edge overlap. The phase diagram of $\tau$ for varying $\epsilon$ and \Gr is shown in Fig. 3E of the main text.

\subsection{Mean-field scaling of the Kuramoto parameter}

In this section we present a mean-field scaling argument for the steady-state order parameter $\langle k \rangle$ as a function of coupling strength $\epsilon$, in the regime $\epsilon \gg \sigma$. Although derived for all-to-all coupling, we argue below that its central prediction — a $1/\epsilon^2$ scaling of $1-\langle k\rangle$ — extends, up to a geometry-dependent prefactor, to the short-range (2D nearest-neighbor) coupling used in our simulations, and we show that it is consistent with observations even outside its strict range of validity ($\epsilon \gtrsim \sigma$).

\paragraph*{Starting point.}
We start from Kuramoto's exact self-consistency condition for the steady, phase-locked branch of the mean-field model, as derived by Kuramoto and reviewed in Strogatz \cite{strogatz2000kuramoto}. Here $g(\omega)$ denotes the probability density from which the intrinsic frequencies $\omega_i$ are drawn, assumed even and unimodal about $\omega=0$; in our case $g$ is a Gaussian of standard deviation $\sigma$. Writing the order parameter as $r=\langle k \rangle$ and the coupling as $K=\epsilon$, the locked oscillators satisfy $\omega = \epsilon\langle k \rangle \sin\theta$ (Strogatz, Eq.~(4.2)), and imposing self-consistency of $\langle k \rangle$ with this locked population gives (Strogatz, Eq.~(4.4)):
\begin{equation}
\langle k \rangle = \epsilon \langle k \rangle \int_{-\pi/2}^{\pi/2} \cos^2\theta\; g\big(\epsilon \langle k \rangle \sin\theta\big)\, d\theta .
\end{equation}
Changing variables to $x=\epsilon\langle k \rangle\sin\theta$ (so that $x$ runs directly over the locked natural frequencies, $|x|\le \epsilon\langle k \rangle$), this becomes the equivalent, exact relation
\begin{equation}
\langle k \rangle = \int_{-\epsilon\langle k \rangle}^{\epsilon\langle k \rangle} \sqrt{1-\frac{x^2}{(\epsilon\langle k \rangle)^2}}\; g(x)\, dx .
\label{eq:selfconsistency}
\end{equation}
This is valid for any even, unimodal frequency distribution $g(\omega)$ and any $\epsilon$ on the partially synchronized branch, i.e. the nontrivial solution $\langle k\rangle>0$ of Eq.~\eqref{eq:selfconsistency} that bifurcates continuously from the incoherent state $\langle k\rangle=0$ at $\epsilon=\epsilon_c=2/(\pi g(0))$ and exists for all $\epsilon\ge\epsilon_c$ \cite{strogatz2000kuramoto}.

\paragraph*{Deep-locked expansion ($\epsilon\gg\sigma$).}
We now specialize to a Gaussian $g(\omega)=(\sigma\sqrt{2\pi})^{-1}e^{-\omega^2/2\sigma^2}$ and take $\sigma\ll\epsilon$. In this regime the locking half-width $\epsilon\langle k \rangle$ is much larger than $\sigma$, so (i) the fraction of the population outside the locking window is exponentially small in $(\epsilon/\sigma)^2$ and can be neglected, and (ii) within the window, $x/(\epsilon\langle k \rangle)\ll1$ over essentially the whole support of $g$, allowing us to Taylor-expand the square root in Eq.~\eqref{eq:selfconsistency}.
Extending the integration limits to $\pm\infty$ (valid up to the exponentially small correction above) and using the Gaussian second moment $\int x^2 g(x)\,dx=\sigma^2$, Eq.~\eqref{eq:selfconsistency} reduces to
\begin{equation}
\label{eq:k_mean_field}
\langle k \rangle \approx 1 - \frac{\sigma^2}{2\epsilon^2\langle k \rangle^2}.
\end{equation}
Since $\langle k \rangle\to1$ at leading order in this limit, we replace $\langle k \rangle^2\to1$ on the right-hand side, consistent with the order of the expansion, giving $1-\langle k \rangle \;\approx\; \sigma^2/2\epsilon^2$.

\paragraph*{Relevance beyond mean field.}
Although the derivation above assumes mean-field, all-to-all coupling, whereas our simulations use short-range coupling on a 2D lattice, we argue that it remains a useful guide to the observed scaling. In short-range Kuramoto-type models, the same torque-balance intuition applies locally: each oscillator's phase deviation is set by a balance between the coupling, which acts through the network's graph Laplacian rather than a single global field, and the local disorder in natural frequencies. For thin-tailed frequency distributions such as the Gaussian used here, this local balance has been shown to produce the same $1-\langle k \rangle \propto \sigma^2/\epsilon^2$ functional form as the mean-field case, with the coupling topology and dimensionality entering only through a multiplicative prefactor \cite{hong2005collective,restrepo2005onset}. We therefore expect that the mean-field calculation above fixes the correct scaling exponent for our system, while the prefactor itself is not expected to match the mean-field value quantitatively.

\paragraph*{Comparison to simulations.}
Fig.~3D of the main text plots $(1-\langle k \rangle)\epsilon^2/\sigma^2$, on a logarithmic scale, as a function of the growth rate \Gr, for several values of $\epsilon$ restricted to $\epsilon>\sigma$ (the regime for which the above asymptotic derivation is expected to apply). At low \Gr~— where \Gr~is not yet large enough to drive the system toward synchronization — the curves collapse onto a plateau of order unity, consistent with the leading-order prediction of Eq.~\ref{eq:k_mean_field}. As \Gr~increases, all curves decrease together, tracking one another closely across the full range of $\epsilon$ shown (different shades of orange), even though the absolute value of $(1-\langle k \rangle)\epsilon^2/\sigma^2$ departs from its asymptotic plateau. This indicates that the $1/\epsilon^2$ scaling itself remains robust throughout, while \Gr~modulates only the overall prefactor — precisely as expected, since larger \Gr~drives the system further into the phase-locked regime for which our derivation was constructed, and where its underlying assumptions become increasingly well satisfied. In this sense, although the calculation is strictly asymptotic ($\epsilon\gg\sigma$, mean-field), it captures the qualitative and, over most of the accessible range of $\epsilon$ and \Gr, the quantitative behavior seen in practice — for instance the collapse is already clearly visible at $\epsilon=0.3$, $\sigma=0.2$, i.e.\ well before the strict $\epsilon\gg\sigma$ limit is formally reached. This agreement, obtained without invoking system-size or lattice-geometry dependence explicitly, supports treating the mean-field result as a leading-order scaling ansatz for our finite, short-range system, rather than as a quantitative prediction from first principles.

\section{Quasi-analytic description of the 1D bundle dynamics}

In this section, we derive a quasi-analytic description of the one-dimensional dynamics in order to explain why the drift increases with the growth rate $v_p$ and decreases with the wavelength $\lambda$. To simplify the notations and not confuse the particle velocity with the growth rate, we refer to the growth rate as $\gamma$ instead of $v_p$ in this section. All distances are rescaled by $\ell_0$, the typical interparticle distance. 

\begin{figure}
    \centering
    \includegraphics[width=0.99\linewidth]{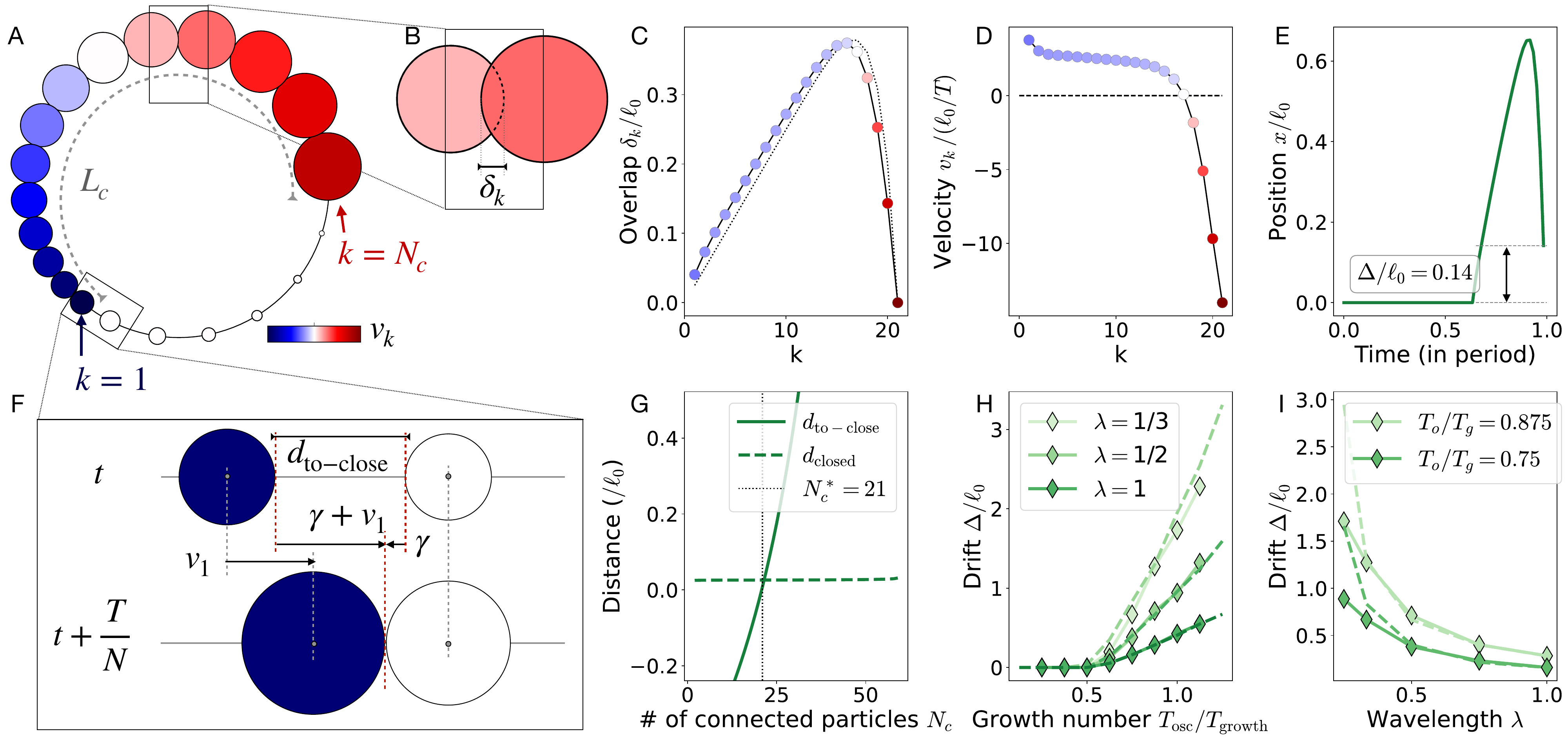}
    \caption{\textbf{The net particle drift increases with the growth number and decreases with the wavelength in a semi-analytical model.}
    (A)~In a one-dimensional chain of particles, only a fraction are in contact, and they cover a length $L_c$. Color codes for the particles' velocity: red is clockwise direction, and blue is trigonometric direction.
    (B)~Neighboring particles $k$ and $k+1$ overlap of a distance $\delta_k$. 
    (C)The overlap can be solved numerically for any set of particles. Points and solid line are a resolution of the system for $20$ particles. Color codes for the particles' velocity (colorbar of (A)). The profile can be approached analytically by Eq.~\ref{eq:profile_fit} (dotted line). 
    (D)~The velocity profile resulting from the overlap profile reveals asymmetric forward and backward displacement of the particles, resulting in a net drift after a period (E). Color codes for the particles' velocity (colorbar of (A)). 
    (F)~Images of the last particle which entered the bundle, and the first particle to enter next at two different times. The time they need to be in contact is set by their initial distance $d_{\mathrm{to-close}}$, their growth rate $\gamma$ and the velocity of the bundle particle $v_1$.
    (G)~This time sets the steady on the number of particle in contact $N_c$. 
    (H--I)~Both $N_c$ and the velocity profile depend on the growth number and the wavelength, confirmed in the ansatz estimate of the overlap profile (dashed lines) or from the explicit numerical solving (diamond symbols).
    }
    \label{fig:1Dmodel}
\end{figure}

We consider a one-dimensional chain of particles with overdamped dynamics. A total of $N$ particles are distributed along the system, while $N_c$ consecutive particles form a connected bundle (see Fig.~\ref{fig:1Dmodel}A). The overlap between neighboring particles  $k$ and $k+1$, drawn in Fig.~\ref{fig:1Dmodel}B, is defined as
\begin{equation}
\label{eq:delta_def}
    \delta_k = \sigma_k + \sigma_{k+1} - (x_{k+1}-x_k),
\end{equation}
where $\sigma_k$ denotes the particle size and $x_k$ its position.

The particle sizes follow the imposed traveling growth profile,
\begin{equation}
    \sigma_k = \gamma t + \gamma \frac{T}{N}k,
\end{equation}
where $T$ is the oscillation period. The wavelength $\lambda$ effectively rescales the number of particles involved in one period of the dynamics. In practice, decreasing $\lambda$ corresponds to solving the same dynamics over a proportionally smaller effective system size.

Neighboring particles interact through a harmonic repulsion,
\begin{equation}
    f_k = U_0 \delta_k,
\end{equation}
such that the total force acting on particle $k$ is
\begin{equation}
    F_k = f_{k-1}-f_k.
\end{equation}

The dynamics is overdamped, with a friction coefficient proportional to the particle size, yielding
\begin{equation}
    \dot{x}_k = \frac{1}{\sigma_k}F_k.
\end{equation}

Taking the time derivative of Eq.~\ref{eq:delta_def} leads to the overlap dynamics
\begin{equation}
\label{eq:delta_dynamic}
    \dot{\delta}_k
    =
    2\gamma
    +
    \frac{U_0}{\gamma}
    \left[
    \frac{\delta_{k-1}}{t+ak}
    -
    \delta_k
    \left(
    \frac{1}{t+ak}
    +
    \frac{1}{t+ak+a}
    \right)
    +
    \frac{\delta_{k+1}}{t+ak+a}
    \right],
\end{equation}
with
\begin{equation}
    a = \frac{T}{N}.
\end{equation}

The bundle contains exactly $N_c$ particles in contact. Consequently, the particles at the edges of the bundle do not overlap with particles outside the bundle, which imposes the boundary conditions
\begin{equation}
    \delta_0 = \delta_{N_c}=0.
\end{equation}

During a time interval $T/N$, one particle depolymerizes and exits the bundle. In a steady state, this process is compensated by the entry of a new particle on the opposite side. We, therefore, solve Eq.~\ref{eq:delta_dynamic} over a time interval $T/N$ and relabel particles after each depolymerization event according to
\begin{equation}
    \delta_k(t+T/N)=\delta_{k-1}(t),
    \qquad
    \delta_1(t+T/N)=0.
\end{equation}

Iterating this procedure yields a steady overlap profile shown in Fig.~\ref{fig:1Dmodel}C. The profile is asymmetric, with vanishing overlaps at both edges of the bundle. This asymmetry directly generates a finite drift.

The particle velocities (Fig.~\ref{fig:1Dmodel}D) are obtained from the force balance,
\begin{equation}
    v_k = \dot{x}_k,
\end{equation}
and the total drift (Fig.~\ref{fig:1Dmodel}E) accumulated during one relabeling event is
\begin{equation}
    \Delta(N_c)
    =
    \frac{T}{N}
    \sum_{k=1}^{N_c} v_k.
\end{equation}

Although Eq.~\eqref{eq:delta_dynamic} does not admit a simple closed-form solution for arbitrary $N_c$, the numerical steady profiles are well captured by the empirical form
\begin{equation}
\label{eq:profile_fit}
    \delta_k
    \simeq
    2\gamma \frac{N}{\lambda}
    \left[
    1-\left(\frac{k}{N_c}\right)^{N_c/2}
    \right].
\end{equation}
This expression should be understood as a scaling ansatz rather than an exact solution. It incorporates three key properties of the steady profile. It is shown as a dotted line in Fig.~\ref{fig:1Dmodel}C. This approximation becomes increasingly accurate for sufficiently large $N$.

First, the amplitude is proportional to the growth rate $\gamma$, since overlaps are produced by particle growth at rate $2\gamma$. Second, the amplitude scales as $N/\lambda$, reflecting the fact that decreasing the wavelength increases the phase gradient of the growth profile and therefore increases the local compression. Third, the expression satisfies the boundary condition $\delta_{N_c}=0$ imposed by the definition of the bundle edge.

The exponent $N_c/2$ provides a compact representation of the boundary layer near the exiting edge of the bundle. Indeed, for $k=N_c-m$ with $m\ll N_c$,
\begin{equation}
    \left(\frac{k}{N_c}\right)^{N_c/2}
    =
    \left(1-\frac{m}{N_c}\right)^{N_c/2}
    \simeq
    e^{-m/2}.
\end{equation}
Thus Eq.~\eqref{eq:profile_fit} corresponds to an overlap profile that is nearly constant in the bulk of the bundle and relaxes exponentially over a few particles near the edge. This is consistent with the steady-state solution of Eq.~\eqref{eq:delta_dynamic}, in which growth generates overlaps throughout the bundle while elastic relaxation removes them at the boundaries.

Equation~\ref{eq:profile_fit} highlights the physical origin of the drift. Increasing the growth rate $\gamma$ increases the overlap amplitude throughout the bundle and therefore enhances the drift velocity. Conversely, increasing the wavelength $\lambda$ reduces overlap gradients and decreases the asymmetry of the profile, leading to a smaller drift.

The previous calculations assume a prescribed bundle size $N_c$. To determine the selected value $N_c^\ast$, we use a geometric self-consistency condition.

Two independent constraints determine the distance between the last particle outside the bundle and the first particle inside the bundle:

\begin{itemize}

\item
During the time interval $T/N$, the leftmost particle of the bundle moves with velocity $v_1$. This determines the distance closed by the expansion of the bundle, denoted $d_\mathrm{closed}$ (see Fig.~\ref{fig:1Dmodel}F).
\begin{equation}
    d_\mathrm{closed} = (v_1+2\gamma)\frac{T}{N}
\end{equation}

\item
Because of periodic boundary conditions, the bundle length $L_c$ (defined in Fig.~\ref{fig:1Dmodel}A) fixes the size of the dilute region outside the bundle and therefore the interparticle spacing in this region. $L_c$ depends on the totaly system length $L$ and determines the distance that must be closed for a new particle to enter the bundle, denoted $d_\mathrm{to-close}$.
\begin{equation}
    d_\mathrm{to-close} = \frac{L-L_c(N_c)}{N-N_c}, \quad L_c(N_c) = \sum_{k=1}^{N_c}(\sigma_k - \delta_k)
\end{equation}
Counterintuitively, $d_\mathrm{to-close}$ increases with $N_c$, more particles in the bundle means less space in the free regions, but the particles at the edge are also smaller, meaning larger distance between them, and the total overlap also increases with $N_c$.

\end{itemize}

For given parameters $(\gamma,\lambda,N,U_0)$, we compute both quantities as functions of $N_c$. The selected bundle size $N_c^\ast$ is obtained from the self-consistency condition
\begin{equation}
    d_\mathrm{closed}(N_c^\ast)
    =
    d_\mathrm{to-close}(N_c^\ast).
\end{equation}
We show an example of the dependency of $d_\mathrm{closed}$ and $ d_\mathrm{to-close}$ with $N_c$ in Fig.~\ref{fig:1Dmodel}G.

We solve this system of equations numerically---distinct from the particle-based simulations of the main text, as it relies only on the reduced overlap dynamics (Eq.~\ref{eq:delta_dynamic}) and the self-consistency condition for $N_c^\ast$---and confirm that the resulting drift increases with the growth number $T_\mathrm{osc}/T_\mathrm{growth}$ (directly proportional to $\gamma$) and decreases with $\lambda$, consistent with the trends reported in Fig.~\ref{fig:1Dmodel}H--I. While the coupling between $N_c^\ast$ and $(\gamma,\lambda)$ prevents a fully closed-form expression for $\Delta$, the scaling ansatz of Eq.~\eqref{eq:profile_fit} makes the origin of these trends transparent: growth injects overlap at a rate proportional to $\gamma$, and this overlap accumulates over a phase gradient set by $\lambda$, so that both larger $\gamma$ and smaller $\lambda$ sharpen the asymmetry of the overlap profile and thus the net drift.

\section{Supplementary data}

\paragraph*{\textbf{Supplementary video S1: Artificial centrosome (AurkA beads) tracking reveals correlation between cytoplasmic flow and cell cycle wave through microtubule dynamics.}} 
Video of a typical experiment with cycling cytoplasmic extract supplemented with AurkA beads as artificial centrosomes. Large field of view is shown via stitching of 4 tiles (10x magnification, 10\% overlap) (on the left) and the top right tile enlarged (on the right). Tracks with dragontails (backward in time - 30 min) are visualized for the same videos. Scale bars $500~\mu m$. 

\paragraph*{\textbf{Supplementary video S2: Phase field computation and artificial centrosome tracking show alignment between wave direction and flows.}} Juxtaposition of the raw microtubule intensity and Aurka beads tracked trajectories (left), with the result of the analysis: background color is the cycling phase field, black arrows illustrate the direction of the phase gradient, and green arrow illustrate the direction of the beads (middle). The histogram (right) counts the occurrence of binned angle difference between the two directional fields. 

\paragraph*{\textbf{Supplementary video S3: Particle simulation shows alignment between wave direction and flows. }} Simulation video illustrates how the individual particles move (left), while the phase field equilibrates to a few sinks and sources (middle). The particles direction (green arrows) is correlated with the phase field gradient (black arrows) and counted in the histogram (right) at each step. 


\end{document}